\documentclass[twocolumn,final,aps,prb,showpacs,superscriptaddress,amsmath,amssymb,amsfonts,floatfix]{revtex4-2}
\usepackage{epsfig}
\usepackage[dvipsnames,usenames]{color}
\usepackage{xcolor}
\usepackage{ulem}
\usepackage{amsmath}
\usepackage{comment}
\usepackage{array}
\usepackage{multirow}
\usepackage{tabularx}
\usepackage{cancel}
\usepackage{float}
\usepackage[colorlinks=true,bookmarks=true, pdfborder={0 0 0},linkcolor=blue,urlcolor=blue,     breaklinks=true,bookmarksnumbered=true]{hyperref}

\emergencystretch=\maxdimen
\begin{document}

\title{Superconductivity in imbalanced bilayer Hubbard model: enhanced $d$-wave and weakened $s^\pm$-wave pairing}
\author{Ziying Jia}
\affiliation{School of Physical Science and Technology, Soochow University, Suzhou 215006, China}

\author{Xun Liu}
\affiliation{School of Physical Science and Technology, Soochow University, Suzhou 215006, China}

\author{Mi Jiang}
\affiliation{School of Physical Science and Technology, Soochow University, Suzhou 215006, China}
\affiliation{Jiangsu Key Laboratory of Frontier Material Physics and Devices, Soochow University, Suzhou 215006, China}

\begin{abstract}
We investigate the bilayer model with two layers of imbalanced densities coupled by the interlayer hybridization. Using the large-scale dynamical cluster quantum Monte Carlo simulation, we discovered that increased hybridization induces a transition in the superconducting pairing from $d$-wave to $s^{\pm}$-wave and the superconducting $T_c$ of $d$-wave pairing exhibits a non-monotonic dependence on the density imbalance. Remarkably, the optimal superconductivity(SC) occurs at a moderate imbalance. Our results support the possibility of $T_c$ enhancement in composite picture where the underdoped layer provides the pairing strength while the overdoped layer promotes the phase coherence. In addition, the SC can be possibly hosted by a single layer, which is reminiscent of our recent exploration on the trilayer Hubbard model. Our present study thus provides new insight that the SC can be enhanced via the layer differentiation.
\end{abstract}

\maketitle

\section{Introduction}
The two-dimensional bilayer Hubbard model has been extensively investigated for their superconductivity (SC), exciton condensation, magnetic properties etc. ~\cite{bilayer1,bilayer2,bilayer3,bilayer4,bilayer5,bilayer6,bilayer7}. Early studies revealed the regulatory effect of charge imbalance on interlayer magnetic coupling~\cite{1999magnetic}. The phase transition between antiferromagnetic and band insulator has also been systematically studied by quantum Monte Carlo (QMC) and other advanced many-body methods\cite{euverte_magnetic_2013,ruger_phase_2014}. With the discovery of high-temperature SC in nickelates\cite{327}, the current research focuses on various pairing instabilities and mechanism in the two-orbital bilayer model\cite{lin_magnetic_2024,ZYY,LJX,maier327,yang_2024,wang_2024,lu_2024,luo_2024,botzel_theory_2024,zhang_trends_2023,yang_possible_2023,liao_electron_2023,botana_2024}. Besides, electron-hole pairs and exciton condensation in bilayer system with tunable interlayer attraction or repulsion is also a widely studied problem\cite{rademaker_2013,eisenstein_2004,bilayer6,huang_biexciton_2020}.

In high-temperature superconductors, the interlayer coupling in bilayer systems is crucial for the pairing symmetry. Previous studies have demonstrated that weakly coupled bilayer systems exhibit $d$-wave pairing \cite{bilayer1, bilayer2, bilayer3}; while stronger interlayer coupling induces a transition from $d$-wave to $s^{\pm}$-wave pairing \cite{maier2011,maier2016}, along with the emergence of additional phases such as the pseudogap \cite{lx2025}. Recent studies have revealed competition between $d$ and $d+is$ pairing in two-dimensional lattice models\cite{pan_competition_2024}, which is governed by carrier density and intralayer hopping. The bilayer systems exhibit similar complexity and early studies demonstrated that ``critical'' interlayer hopping could potentially stabilize the $d+is$ pairing symmetry\cite{maier2011}.
Notably, ultracold atomic experiments observing competing magnetic orders~\cite{bilayer4,bohrdt_2021}, non-local density fluctuation \cite{cold2} and superfluid \cite{cold3} confirmed that the bilayer model has tangible physical relevance.

Apart from the standard situation with two equivalent layers, the bilayer systems with two imbalanced layers are of interest as well. For example, the composite picture provides the possibility that an underdoped layer (UL)  enhances the pairing scale while the other optimally or overdoped layer (OL) supplies phase stiffness~\cite{Kivelson,Kivelson2}. This concept has generated considerable interest in understanding how multilayer systems can enhance the superconducting critical temperature $T_c$. For instance, enhanced SC has been observed in superlattice systems with repulsive Coulomb interaction  \cite{maier2008,fontenele_increasing_2024}, as well as in composite superconductor/metal bilayer with attractive Coulomb interactions \cite{maier2022,zhang_optimizing_2025}, where $T_c$ exceeds that of single-layer systems \cite{maier2008}. Similarly, prior work on composite superconductors suggested that minimizing the thickness of the metallic layer can suppress phase fluctuations, thereby maximizing $T_c$~\cite{Dror}.

\begin{figure}[t!]
\psfig{figure=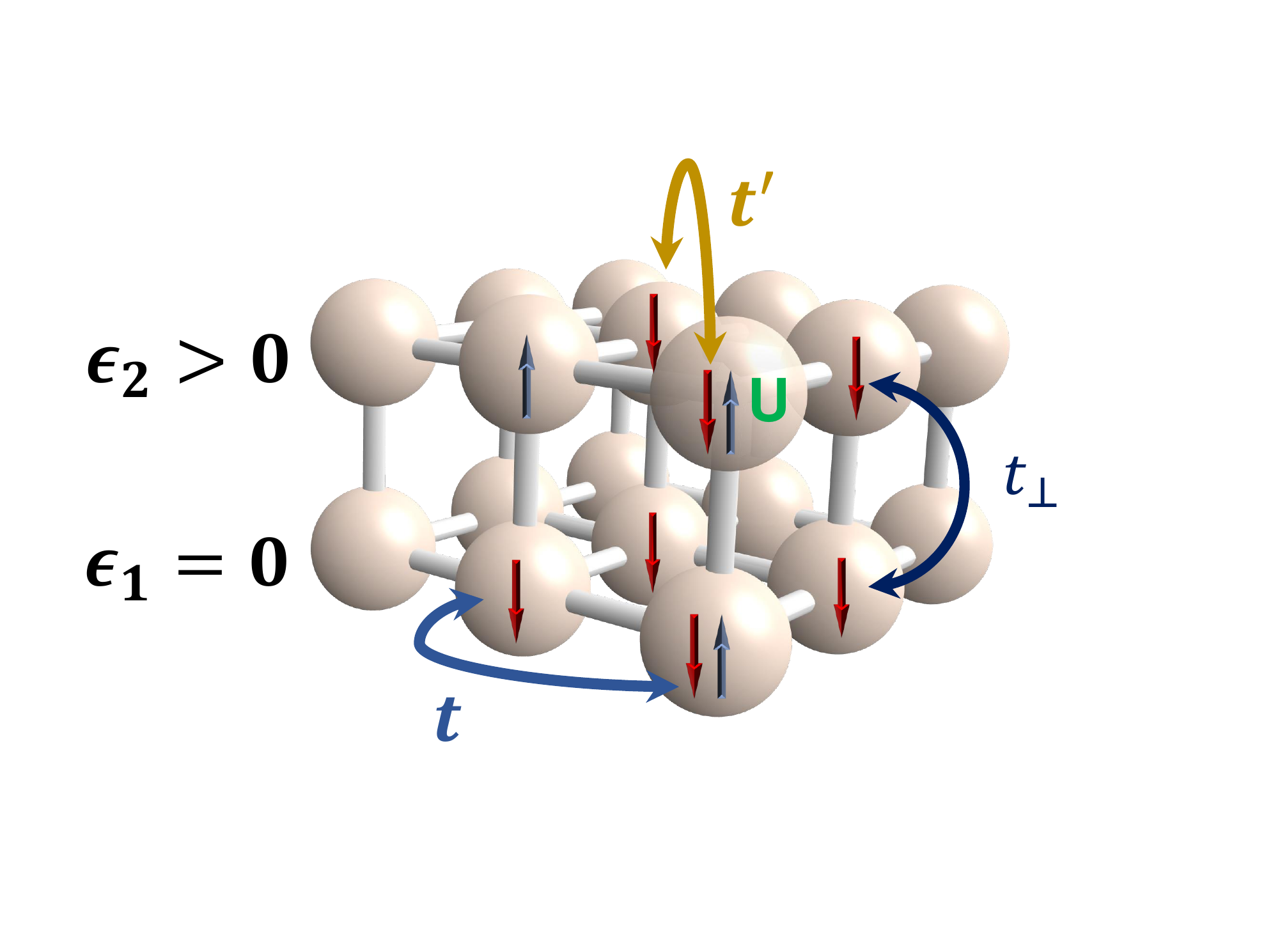, height=5cm,width = .53\textwidth,trim=0cm 4.5cm 0cm 3cm,clip=true}
\caption{Schematic diagram of the imbalanced bilayer Hubbard model. The blue and yellow arrows represent nearest-neighbor and next-nearest-neighbor hoppings respectively and the dark arrow indicates the interlayer hybridization. The onsite energy $\epsilon_2>0$ is introduced to tune the density imbalance between layers.}
\label{lattice}
\end{figure}

Motivated by these findings, we investigate the bilayer Hubbard model with repulsive onsite interactions to explore how imbalanced electron densities across the two layers influence the SC properties. Specifically, we not only map out the transition between $d$-wave and $s^{\pm}$-wave pairing symmetries at weak-to-intermediate hybridization strengths but also identify higher $T_c$ in systems with imbalanced electron densities. We systematically analyze the relationship between $T_c$ and electron density distribution, as well as the SC phase transition driven by interlayer hybridization. 
This paper is organized as follows: Section II introduces the model and methodology; Section III presents the phase diagram and associated properties of $s^\pm$-wave and $d$-wave pairings; and Section IV provides the summary and outlook.

\begin{figure}[t!]
\psfig{figure=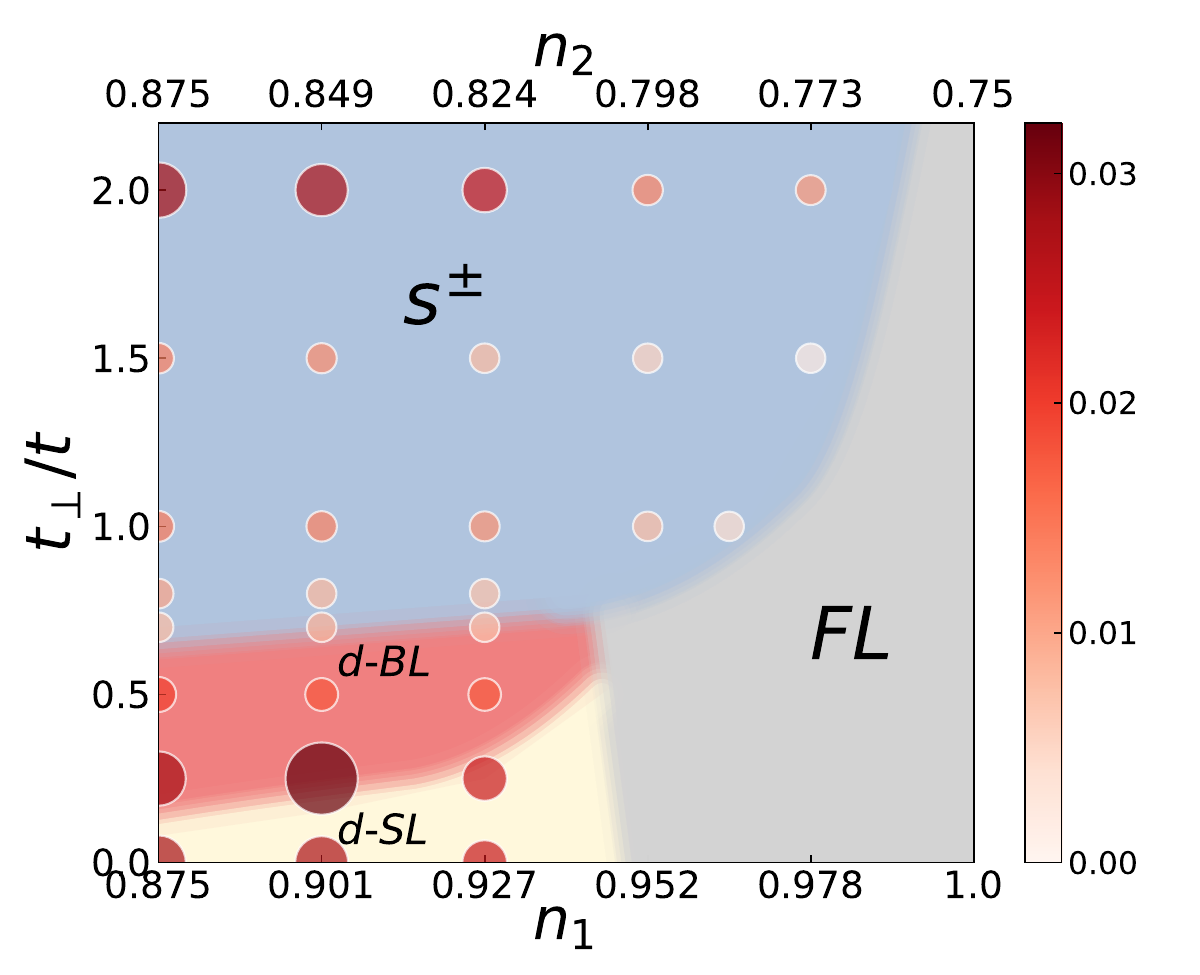, height=7cm,width = .5\textwidth, clip}
\caption{The density versus interlayer hybridization phase diagram of imbalanced bilayer Hubbard model consists of Fermi liquid (FL), $d$-wave on both layers ($d$-BL), $d$-wave on single layer ($d$-SL) with lower doping, and $s^{\pm}$-wave SC phases. The estimated phase boundaries are denoted by dashed green lines. The magnitude of $T_c$ is indicated by the size and color of circles. The fixed average electron density is $n=0.875$.}
\label{phase}
\end{figure}

\section{Model and Method}
\subsection{Imbalanced Bilayer Hubbard model}
We consider the bilayer Hubbard model on a two-dimensional square lattice as shown in Fig.~\ref{lattice} with the Hamiltonian
\begin{align}
    H =& -t \sum_{\langle i j \rangle \sigma m} c_{i m \sigma}^\dagger c_{j m \sigma}  - t'\sum_{\langle\langle i j \rangle\rangle \sigma m} \notag c_{i m \sigma}^\dagger c_{j m \sigma} \\ \notag
     &- t_\perp \sum_{i \sigma} c_{i 1 \sigma}^\dagger c_{i 2 \sigma}+\mathrm{H}.c. \\
      &+U \sum_{i m} n_{i m \uparrow} n_{i m \downarrow} + \sum_{i \sigma m} (\epsilon_{2} \delta_{m 2} - \mu) n_{i m \sigma} \notag
\end{align}
where $c_{i \sigma}^{\dagger}$ ($c_{i \sigma})$ creates (annihilates) an electron with spin $\sigma$ (=$\uparrow$,$\downarrow$) in $m$-th(=1, 2) layer at site $i$. We set the intralayer nearest-neighbor hopping $t=1$ as the unit of energy. The conventional parameters for the intralayer next-nearest-neighbor hopping $t'=-0.15t$ and on-site Coulomb interaction $U=7t$ are adopted \cite{Maier2019, lx2025}. The chemical potential $\mu$ fixes the average density, which is chosen at a characteristic $n=0.875$ throughout the present work for simplicity. Other average densities might not induce significant difference from our presented results below. Two important parameters include the additional site energy ($\epsilon_2$) to adjust the electron density imbalance as well as the tunable interlayer hybridization $t_\perp$. Notably, the density distribution remains nearly unaffected by the hybridization (see Appendix Fig.~\ref{density}).

\subsection{Dynamical cluster approximation (DCA)}
DCA with the continuous-time auxiliary-field (CT-AUX) quantum Monte Carlo (QMC) cluster solver~\cite{Hettler98,Maier05,code,GullCTAUX} is employed to numerically solve the bilayer model.
As a celebrated quantum many-body numerical method, the DCA evaluates various physical observables in the thermodynamic limit by mapping the bulk lattice problem onto a finite cluster embedded in a mean-field bath in a self-consistent manner~\cite{Hettler98,Maier05}, which is realized by the convergence between the cluster and coarse-grained (averaged over a patch of the Brillouin zone around a specific cluster momentum $\mathbf{K}$) single-particle Green's functions. 
In particular, the short-range interactions within the cluster are treated exactly with various numerical techniques, e.g. CT-AUX used in the present study; while the longer-ranged physics is approximated by a mean field hybridized with the cluster. Therefore, increasing the cluster size systematically approaches the exact result in the thermodynamic limit. 
The finite cluster size essentially approximates the whole Brillouin zone by a discrete set of $\mathbf{K}$ points so that the self-energy $\Sigma(\mathbf{K},i\omega_n)$ is a constant function within the patch around a particular $\mathbf{K}$ and a step function in the whole Brillouin zone.
Generically, the quantum embedding methods including DCA have a better minus sign problem than the finite-size QMC simulations because of the hosting mean field. Most of our calculations were conducted with $N_c=16(=8\times2)$-site DCA cluster to incorporate the momentum space resolution including in-plane nodal $\mathbf{K}= (\pi/2,\pi/2)$ and antinodal $\mathbf{K} = (\pi,0)$ directions.

\subsection{Bethe-Salpeter equation (BSE)}

The superconducting properties can be studied by solving the Bethe-Salpeter equation (BSE) in its eigen-equation form in the particle-particle channel~\cite{Maier2006,scalapino2007numerical}
\begin{align} \label{BSE}
    -\frac{T}{N_c}\sum_{K'}
	\Gamma^{pp}(K,K')
	\bar{\chi}_0^{pp}(K')\phi_\alpha(K') =\lambda_\alpha(T) \phi_\alpha(K)
\end{align}
where $\Gamma^{pp}(K,K')$ denotes the irreducible particle-particle vertex of the effective cluster problem, with $K=(\mathbf{K}, i\omega_n)$ combining the cluster momenta $\bf{K}$ and Matsubara frequencies $\omega_n=(2n+1)\pi T$, and $\phi_\alpha(K)$ represents the eigenvector solved from BSE with the $\alpha$-wave pairing symmetry. 

The coarse-grained bare particle-particle susceptibility
\begin{align}\label{eq:chipp}
	\bar{\chi}^{pp}_0(K) = \frac{N_c}{N}\sum_{k'}G(K+k')G(-K-k')
\end{align}
is obtained via the dressed single-particle Green's function,
\begin{align}
G(k)\equiv G({\bf k},i\omega_n) =
[i\omega_n+\mu-\varepsilon_{\bf k}-\Sigma({\bf K},i\omega_n)]^{-1}
\end{align}
where $\mathbf{k}$ belongs to the DCA patch surrounding the cluster momentum $\mathbf{K}$ with the chemical potential $\mu$ and the dispersion relation
\[
\varepsilon_\mathbf{k}
=
\begin{bmatrix}
E_\mathbf{k} & -t_\perp\\
-t_\perp & \epsilon_2+E_\mathbf{k}
\end{bmatrix}
\]
where $E_\mathbf{k}=-2t(\cos k_x+\cos k_y)-4t'\cos k_x\cos k_y$, and $\Sigma({\mathbf K},i\omega_n)$ the cluster self-energy. In practice, we calculate 32 or more discrete points for both the positive and negative fermionic Matsubara frequency $\omega_n=(2n+1)\pi T$ mesh for measuring the two-particle Green's functions and irreducible vertices. Therefore, the BSE Eq.~\eqref{BSE} reduces to an eigenvalue problem of a matrix of size $(64N_c)\times (64N_c)$. 

The corresponding pairing operator is defined as
\begin{align}
\Delta_\alpha^{\dagger} &= \frac{1}{\sqrt{N}} \sum_{\mathbf{k}} g_\alpha^{}(\mathbf{k}) c_{\mathbf{k}}^\dagger c_{\mathbf{-k}}^\dagger
\end{align}
Here, $\alpha$ denotes the pairing symmetry channel and the $s^\pm$-wave and $d$-wave symmetries are represented by the form factors $g_\alpha(\mathbf{k})=\cos{k_z}$, $\cos{k_x} - \cos{k_y}$ respectively.

To systematically investigate the interplay between interlayer hybridization and density imbalance, we first analyze the temperature evolution of the BSE eigenvalue $\lambda_\alpha(T)(\alpha=s^\pm,d )$.
 The observation of BCS-like logarithmic temperature dependence would indicate behavior analogous to that observed in the $d$-wave superconducting phase of the conventional single-band Hubbard model in the overdoped regime~\cite{Maier2019,eigenlog}. Conversely, the emergence of linear or exponential temperature dependence would suggest non-BCS pairing fluctuations, similar to those characteristic of the pseudogap regime in the single-band model~\cite{Maier2019}.


\section{Results}
Our main result is shown in the phase diagram Fig.~\ref{phase}, which consists of the $d$-wave and $s^{\pm}$-wave SC phases controlled by the hybridization $t_\perp$ and electron density imbalance tuned by $\epsilon_2$. 
For clarity, the magnitude of $T_c$ is indicated by both the size and color of the circles.
At large $t_\perp$, the bilayer system is dominated by the interlayer pairing so that $s^{\pm}$-wave until large enough density imbalance finally destroys the SC to result in a Fermi liquid (FL). Interestingly, at small $t_\perp$, apart from the conventional $d$-wave pairing on both layers ($d$-BL), there exists a sizable regime with $d$-wave pairing solely on single layer ($d$-SL), which is reminiscent of our recent work of the trilayer Hubbard model~\cite{d_value2}. Intriguingly, the globally optimal $T_c$ occurs near the boundary between these two regimes.  

In the following subsections, we provide a detailed analysis of both $s^{\pm}$-wave and $d$-wave superconducting pairing under varying hybridization strengths and density imbalance conditions. Unless explicitly noted otherwise, all calculations are performed at a fixed average electron density of $n=0.875$.

\begin{figure}[h!]
\psfig{figure=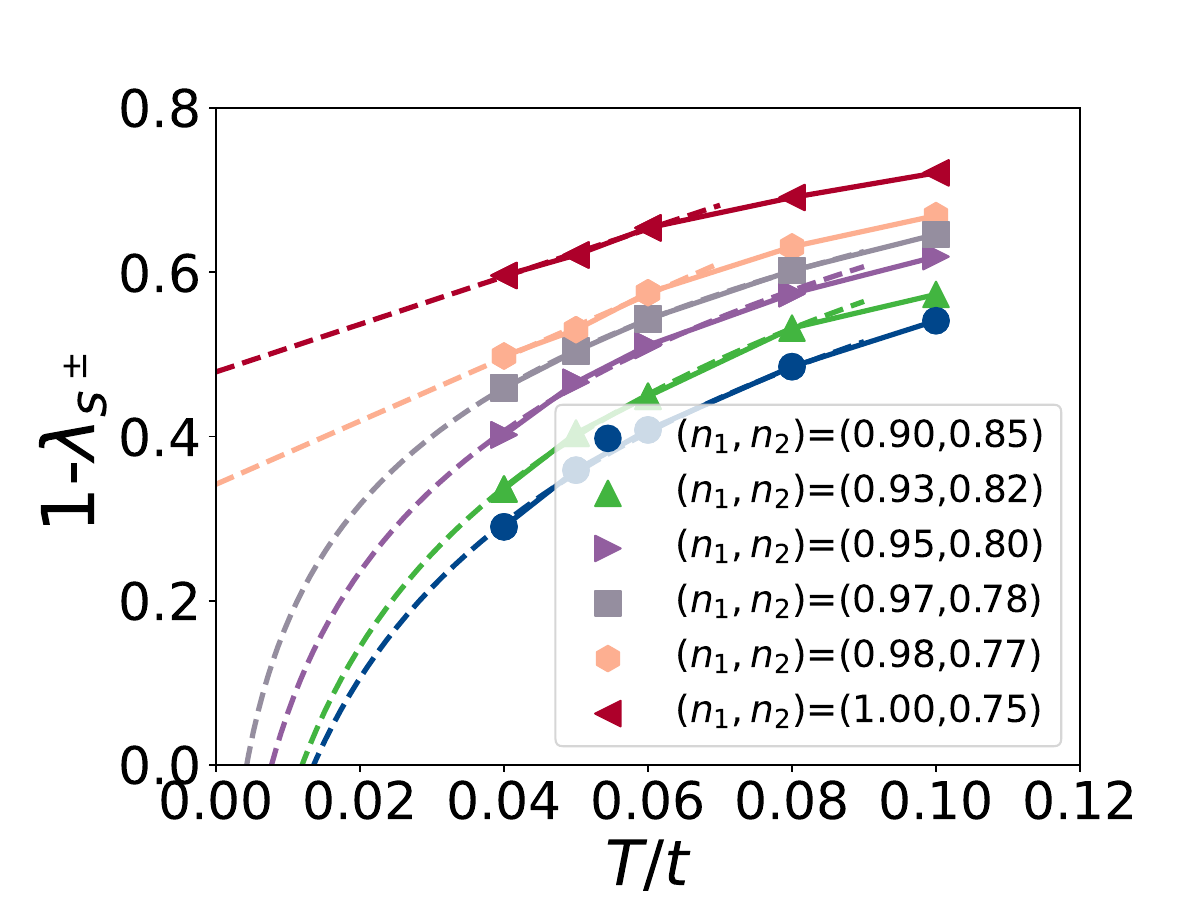, height=5.5cm,width = .45\textwidth,clip}
\caption{Temperature evolution of $1-\lambda_{s^\pm}(T)$ for varying density distribution at the characteristic $t_\perp/t =1.0$.}
\label{slambda}
\end{figure}

\begin{figure}[h!]
\psfig{figure=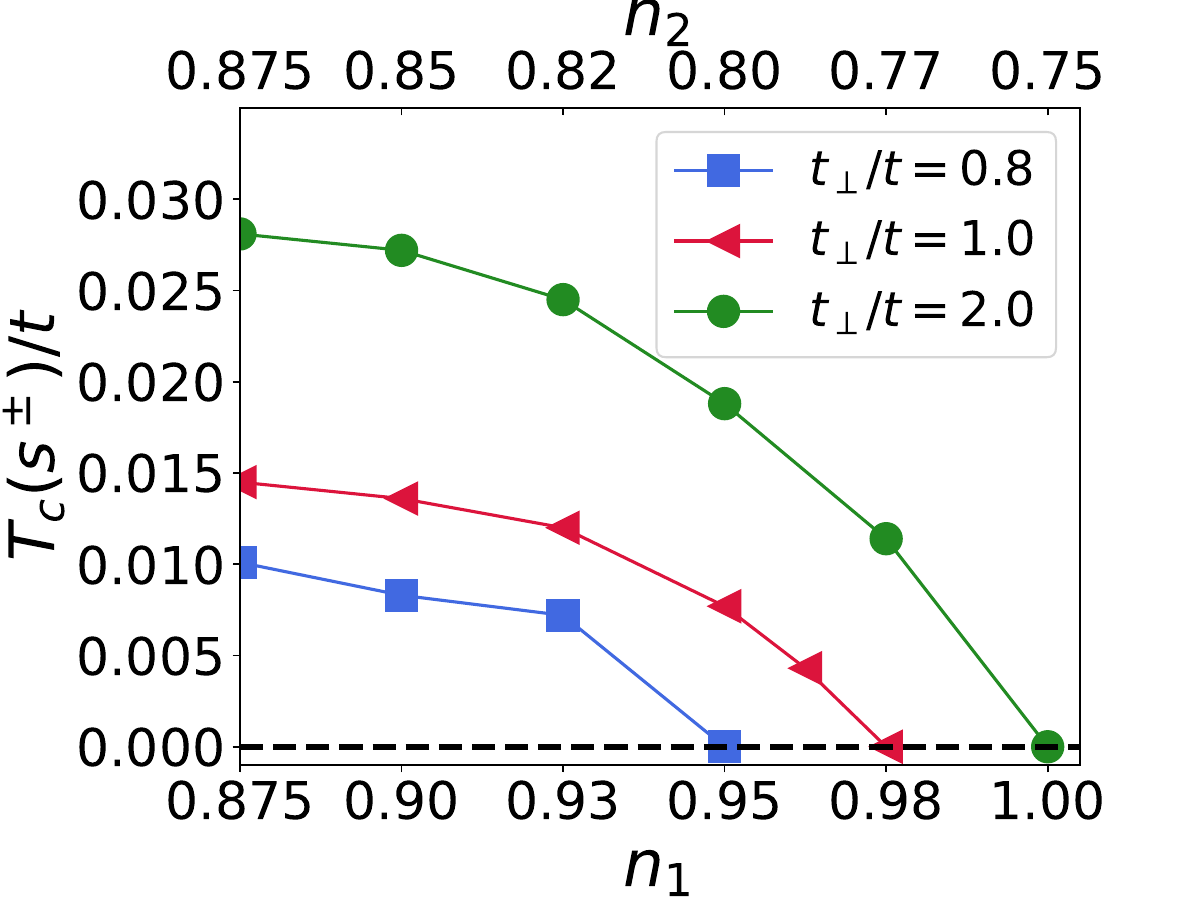, height=5cm,width = .43\textwidth, trim={0 0 0 0}}
\caption{The density evolution of $s^\pm$-wave paring $T_c$ at the three characteristic hybridization $t_\perp/t=0.8,1.0$ and $2.0$.}
\label{Tcs}
\end{figure}

\begin{figure*} [t]
\psfig{figure=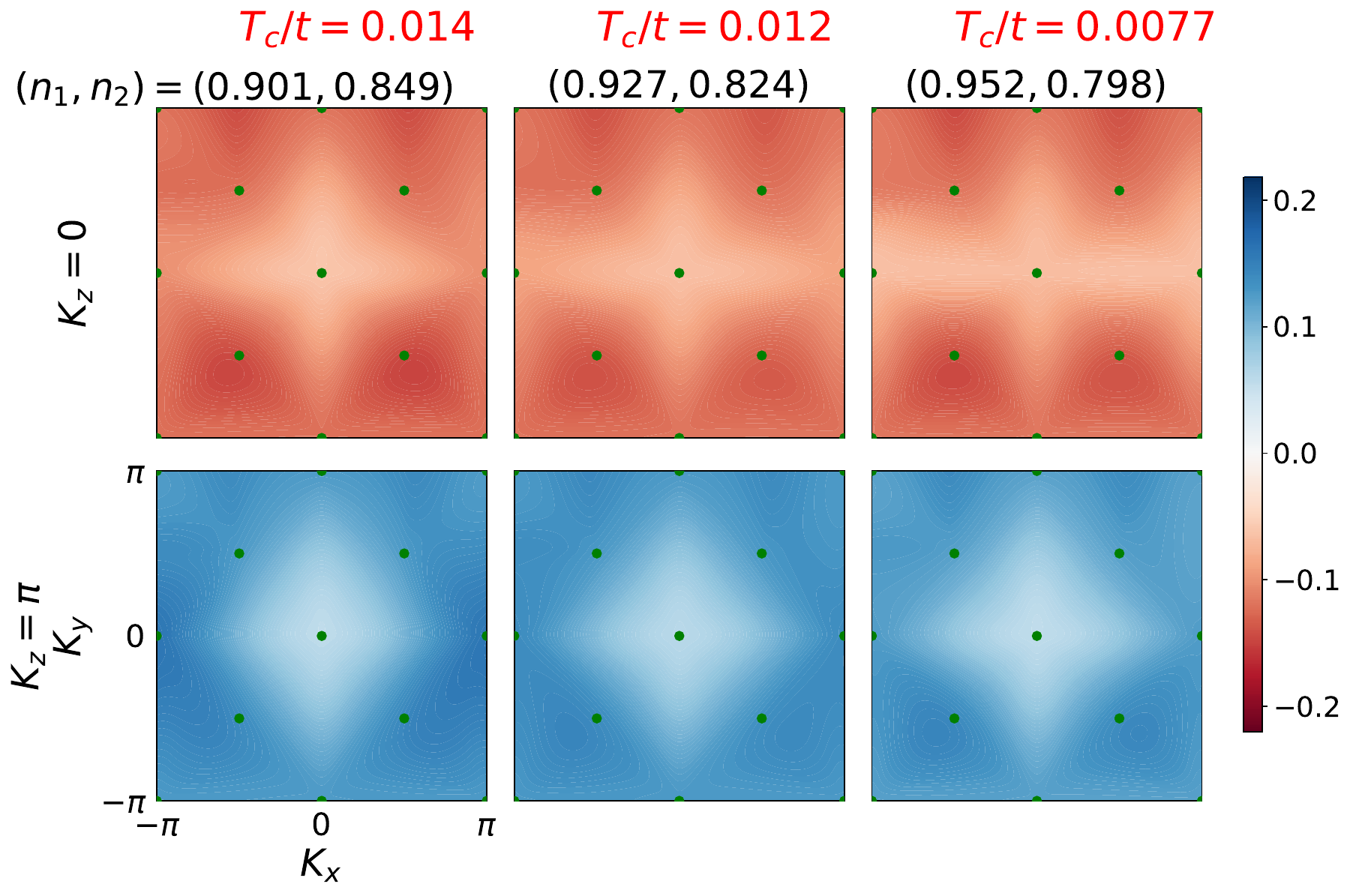, height=10cm,width = 0.98\textwidth,clip}
\caption{The eigenvector $\phi_{s^{\pm}}(\mathbf{K},\pi T)$ of bonding ($\mathbf{K}_z=0$) and anti-bonding ($\mathbf{K}_z=\pi$) at the lowest simulated temperature $T/t = 0.04$ for various density distributions at $t_\perp/t = 1.0$.}
\label{svector}
\end{figure*}

\subsection{$s^{\pm}$-wave pairing at large hybridization}

We first focus on the large $t_\perp$ regime where the $s^{\pm}$-wave pairing dominates. In Fig.~\ref{slambda}, we present the extrapolated $1-\lambda_{s^{\pm}}(T)$ curves for various density distributions at the characteristic hybridization $t_\perp/t=1.0$. Most curves show the BCS-like logarithmic evolution except for the situations when the underdoped layer is close to half-filling, which exhibits the linear temperature dependence instead, implying for the absence of SC as far as one layer approaches to undoped Mott insulator limit. Besides, the pairing instability apparently weakens with the density imbalance monotonically.

The $T_c(s^{\pm})$ is expected to decrease as the density imbalance increases, as validated in Fig.~\ref{Tcs} explicitly. Given that the $s^{\pm}$-wave pairing originates from the interlayer hybridization, the large density imbalance naturally suppresses the pairing. We remark that the eigenvectors are at a fixed temperature above $T_c$ but $\phi_{s^{\pm}}$ still shows the characteristic feature for $s^{\pm}$ pairing in Fig.~\ref{svector} even though the estimated $T_c$ is zero, which can be hinted as the gradual vanishing of $\phi_{s^{\pm}}$ in the anti-bonding band.
Note also that $T_c(s^{\pm})$ is generically lower than its $d$-wave counterpart.

\subsection{$d$-wave pairing with enhanced $T_c$}

\begin{figure}
\psfig{figure=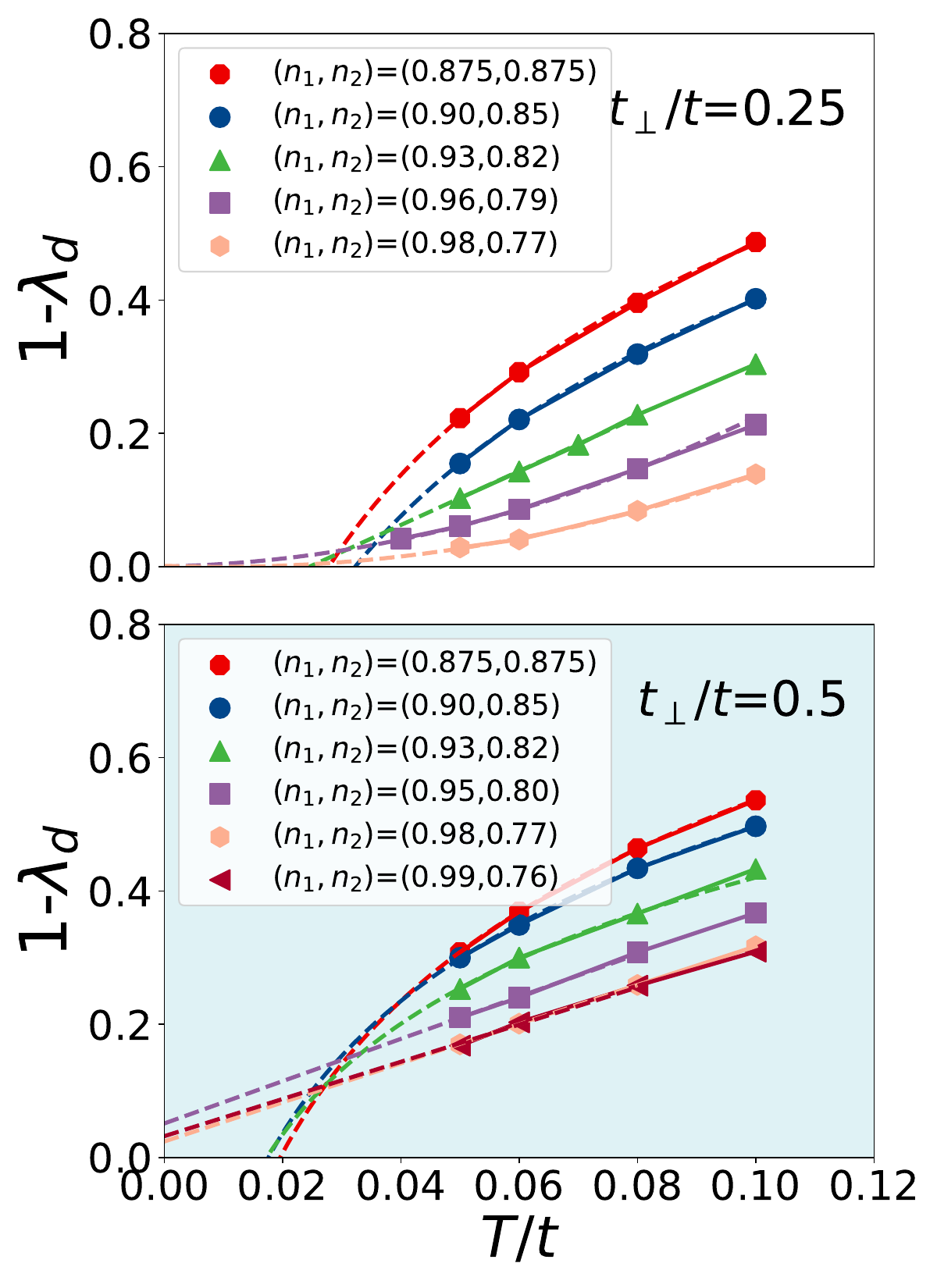, height=11cm,width = .48\textwidth, clip}
\caption{Temperature evolution of $1-\lambda_{d}(T)$ for varying density distribution at two characteristic hybridization $t_\perp/t = 0.25$ and $0.5$. }
\label{dlambda}
\end{figure}

We next switch to the small $t_\perp$ regime where the $d$-wave pairing will dominate over the $s^{\pm}$-wave. The temperature evolution $1-\lambda_d(T)$ curves in Fig.~\ref{dlambda} reveal the density distribution and hybridization dependence distinct from Fig.~\ref{slambda}. Specifically, the moderate density imbalance between the two layers can enhance the $d$-wave $T_c$ at small $t_{\perp}/t=0.25$, but this instability weakens with increasing hybridization strength to larger $t_{\perp}/t=0.5$.

\begin{figure}
\psfig{figure=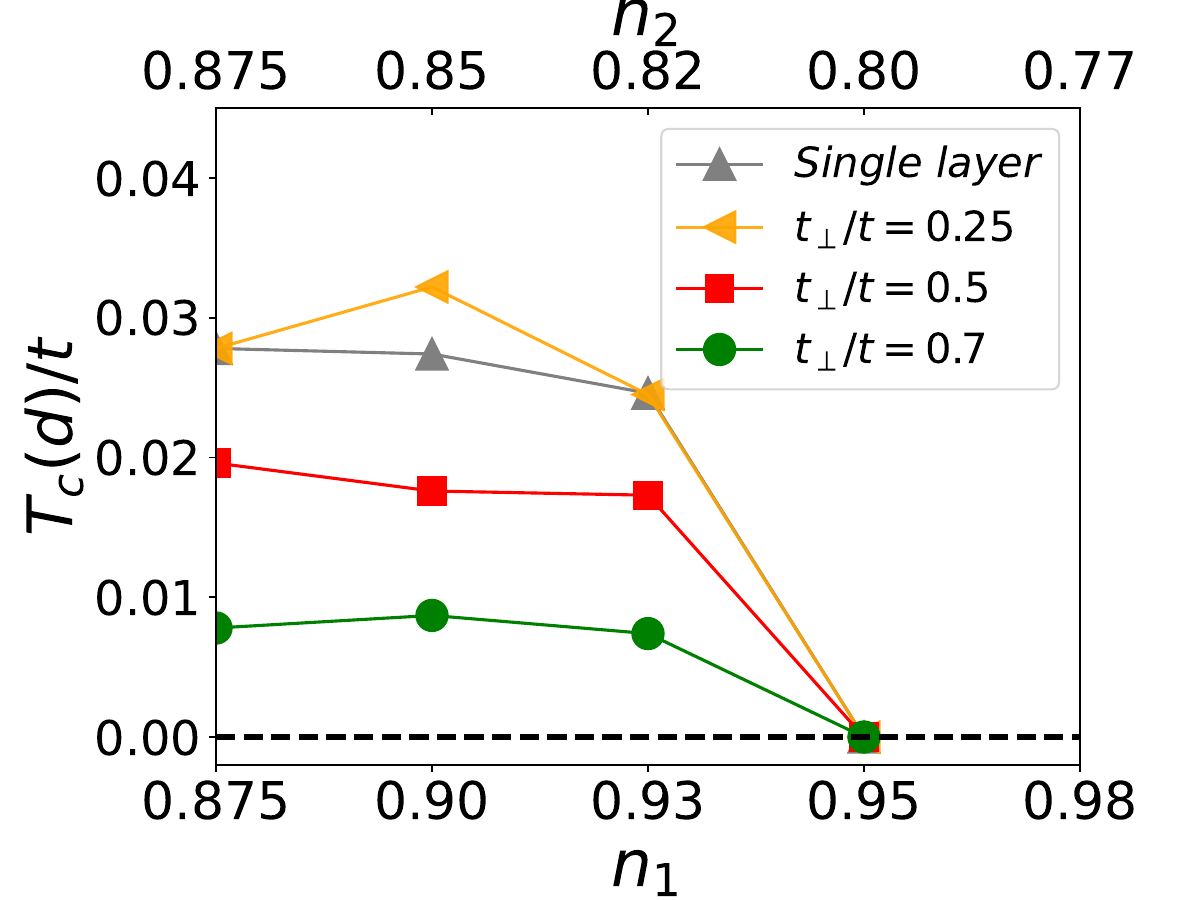, height=6.0cm,width = .48\textwidth, trim={0 0 0 0}}
\caption{Density distribution dependency of $d$-wave paring SC critical temperature $T_c$ at three fixed hybridization $t_\perp/t=0.25,0.5$ and $0.7$ compared with the single layer case.}
\label{Tcd}
\end{figure}

As shown in Fig.~\ref{Tcd}, the $T_c$ for $d$-wave pairing decreases monotonically with increasing hybridization from $t_\perp/t=0.25$ to $0.7$ in all density distributions examined, in contrast to the trend observed for $s^{\pm}$-wave pairing as discussed above.  Most notably, the $T_c$ curves in Fig.~\ref{Tcd} exhibit a non-monotonic dependence on density imbalance before vanishing around $n_1-n_2=0.18$, with this critical value remains unchanged as $t_\perp/t$ grows from 0.25 up to 0.7. Note that the optimal $T_c$, surpassing the density-balanced value, only occurs when the density distribution is around $(n_1,n_2)\sim(0.9,0.85)$ at relatively small $t_{\perp}/t=0.25$. 

\begin{figure*} 
\psfig{figure=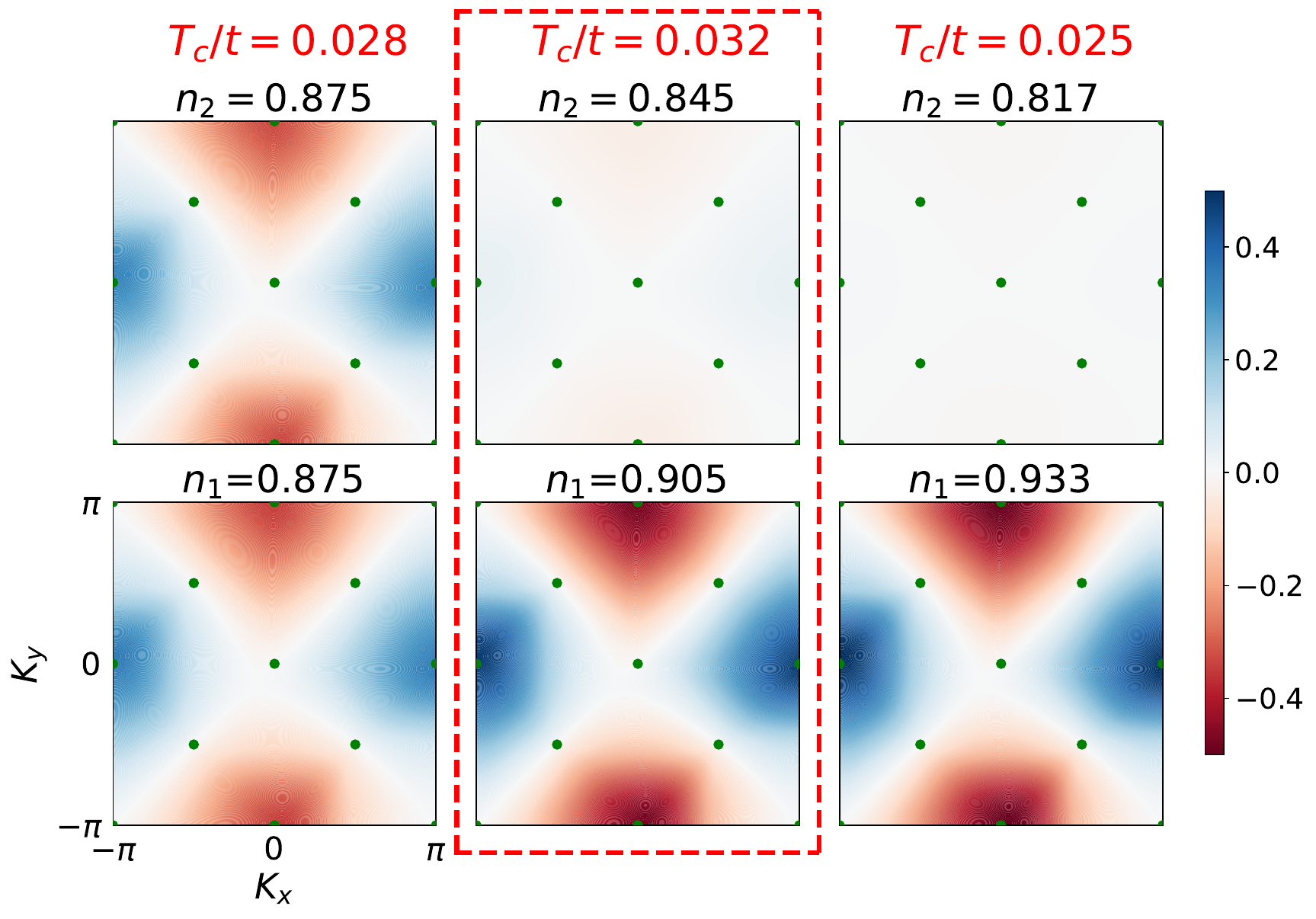, height=10cm,width = 0.95\textwidth,clip}
\caption{The eigenvector $\phi_d(\mathbf{K},T)$ of the OL (upper) and the UL (lower) at the lowest simulated temperature $T/t = 0.05$ for various density distributions at $t_\perp/t = 0.25$. The red box highlights the density imbalance for the optimal $T_c$.}
\label{dvector}
\end{figure*}

Concurrently, as shown in Fig.~\ref{dvector}, the $d$-wave eigenvector $\phi_d$ in the OL disappears when its density decreases, while the pairing strength between the UL and OL differentiates from each other with larger density imbalance. Therefore, we contend that the mechanism of $T_c$ enhancement in the imbalance bilayer model could be potentially accounted for by the composite picture \cite{Kivelson}, where the UL hosts the large pairing scale and the OL provides the strong phase stiffness. The highest $T_c$ likely originates from the maximal pairing UL combined with the phase stiffness contributions from the OL. Besides, the different feature of eigenvectors on the OL and UL suggests that the $d$-wave pairing SC is divided into two distinct phases, labeled as $d$-BL and $d$-SL in Fig.~\ref{phase}, with the highest $T_c$ achieved around their boundary. According to Fig.~\ref{FL1} in Appendix, it is observed that, with increasing density imbalance $n_1-n_2$, the $d$-wave SC diminishes and eventually transits into the Fermi liquid phase as shown in Fig.~\ref{phase}.

\subsection{SC pairing symmetry transition}

The previous work \cite{maier2011} has demonstrated a superconducting pairing symmetry transition from $d$-wave to $s^\pm$-wave. Similarly, in our imbalanced bilayer model, we observe both $s^\pm$-wave pairing at large hybridization (Fig.~\ref{svector}) and $d$-wave pairing at weak hybridization (Fig.~\ref{dvector}), indicating the existence of such SC pairing transitions.

\begin{figure} 
\psfig{figure=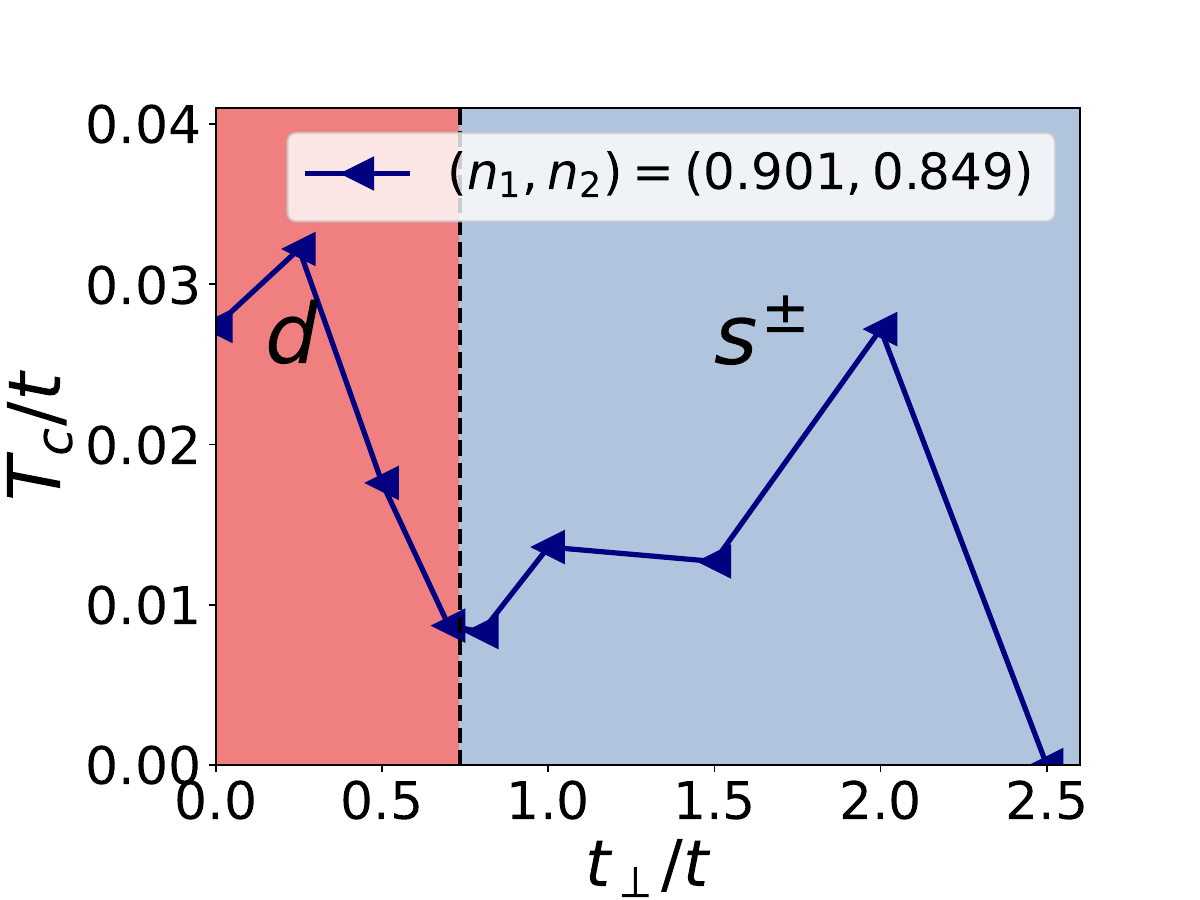, height=6.2cm,width = .48\textwidth, trim={0 0 0 0}}
\caption{The hybridization dependence of $T_c$ at a characteristic density distribution $(n_1,n_2)=(0.901,0.849)$. The $d$-wave pairing and $s^{\pm}$-wave pairing are divided by the dashed line at $t_\perp/t=0.8$.}
\label{d_s}
\end{figure}

Fig.~\ref{d_s} illustrates the non-monotonic dependence of the extracted $T_c$ on the hybridization strength at a fixed density distribution $(n_1, n_2) = (0.901, 0.849)$. The curve reveals markedly different sensitivities of $d$-wave and $s^\pm$-wave pairing to the hybridization. The analysis of layer-resolved pairing in Fig.~\ref{dvector} has revealed that at small $t_\perp$ values, $d$-wave pairing can occur exclusively in the UL.
However, the optimal $T_c$ is achieved when a strengthened $d$-wave pairing in the UL  coincides with the emergence of a nascent $d$-wave state in the OL as shown in middle panel in Fig.~\ref{dvector}.
In contrast, the larger interlayer hybridization induces $d$-wave pairing in both layers ($d$-BL region). While $s^\pm$-wave pairing gradually enhances at intermediate hybridization ($t_\perp/t=0.8$) and reached its optimal pairing conditions at $t_\perp/t=2.0$, this instability is gradually destroyed by stronger hybridization, consistent with our previous work on the balanced bilayer model~\cite{lx2025}.  
As shown in Fig.~\ref{phase}, the critical hybridization strength increases with the density difference $(n_1-n_2)$. Importantly, the highest $T_c$ achieved for $d$-wave pairing generally exceeds that of $s^\pm$-wave pairing.

At sufficiently high density imbalance, the system transits into a Fermi liquid state, as indicated by the temperature-dependent zero-frequency Green function and frequency-dependent self-energy, which will be discussed more in Fig.~\ref{FL2} of Appendix.

\section{Summary and outlook}
In summary, we performed the dynamical cluster quantum Monte Carlo simulations to  systematically explore the effects of the interlayer hybridization and density imbalance on superconducting properties in imbalanced bilayer system.  

The mapped rich phase diagram in Fig.~\ref{phase} reveals the competition between $d$-wave and $s^{\pm}$-wave SC. Intriguingly, it is discovered that the density imbalance is beneficial to the $d$-wave SC and can potentially boost $T_c$ for some cases of density imbalance. Moreover, the SC can be solely hosted by one layer with lower hole doping, which is reminiscent of our recent investigation of the trilayer Hubbard model \cite{d_value2}. This suggests that the $T_c$ enhancement can be accomplished via generic density imbalance in a composite layered system so that the composite picture~\cite{Kivelson} provides the plausible framework for interpreting the enhanced $T_c$. In this picture, the other layer with weaker gap opening is essential for the SC. It is also found that the BSE eigenvector of the OL can open a small $d$-wave gap in low temperatures, which might be induced by the UL via proximity effect. Put another way, our observation that the BSE eigenvector does not exhibit significant weight on the OL is not a proof that the superfluid stiffness is 
not established by tunneling into this layer. Other advanced many-body methods are strongly desired for resolving these issues.

For the $s^{\pm}$-wave pairing regime at large hybridization $t_\perp$, the highest $T_c$ emerges at balanced system and the density imbalance is generally detrimental to SC because of the interlayer pairing nature of $s^{\pm}$-wave SC. One interesting extension is to explore if the previously conjectured coexistence of $d$-wave and $s^{\pm}$-wave pairing \cite{maier2011} can be stabilized in the presence of density balance. 

Given that the bilayer model is relevant to various context of physical systems such as real materials and ultracold atomic systems\cite{bilayer4,cold2,cold3,bohrdt_2021,2024Dissipationless}, it is anticipated that the present study on the influence of the density imbalance provides more insight on this well-established model, especially in terms of the $T_c$ enhancement and its connection to the composite picture. It is worthwhile articulating that the ultracold atomic system has realized the bilayer square lattice model experimentally~\cite{bilayer4}. Recent studies~\cite{cold_2025} have shown that such platform can simulate twisted bilayer structures, where interlayer density–density interactions give rise to a variety of states. Based on these developments, we anticipate that such quantum simulators will further enable the exploration of our density-imbalanced bilayer models in future.


\section{Acknowledgement}
We would like to thank Xianxin Wu, Jun Zhan, and Fan Yang for illuminating discussions.
We acknowledge the support by National Natural Science Foundation of China (NSFC) Grant No.~12174278.


\section{APPENDIX}

\subsection{Hybridization independent density distribution}

\begin{figure}[t!]
\psfig{figure=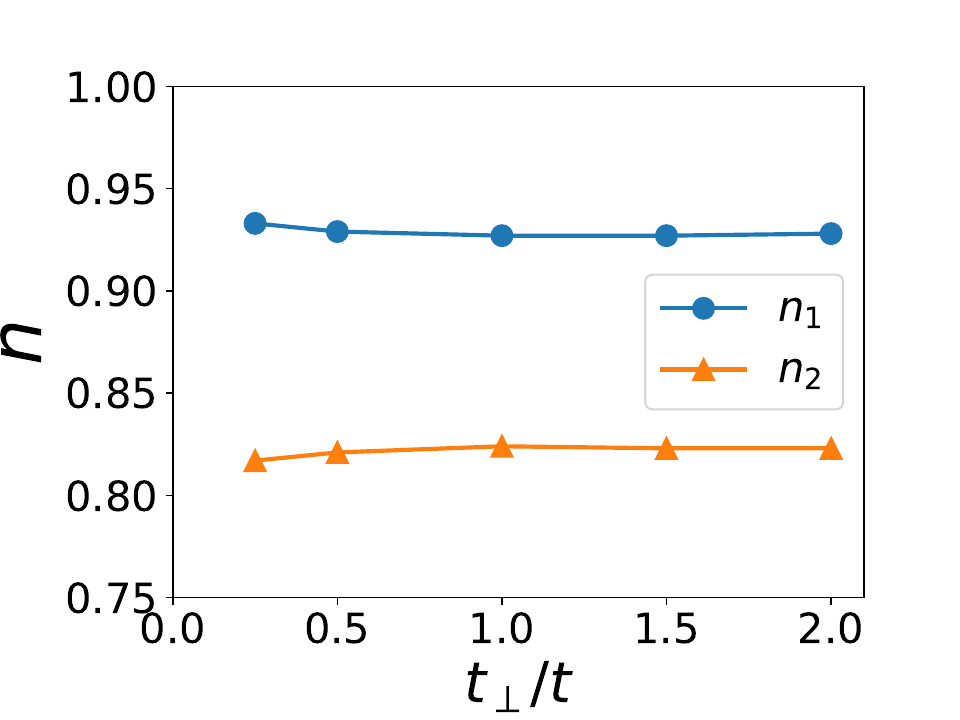, height=6cm,width = .45\textwidth, trim={0 0 0 0}}
\caption{The hybridization dependence of the density distribution at $\epsilon_2/t = 1.0$.}
\label{density}
\end{figure}

Fig.~\ref{density} displays the calculated density distribution for a representative set of parameters ($\epsilon_2=1.0t$ and $t_\perp/t$ varies from 0.25 to 2.0). The density of each layer almost keeps unchanged so that when we use the parameter $\epsilon_2$ to adjust the density distribution in the bilayer model, the influence of $t_\perp$ on the density distribution is negligible. This provides a convenient way to study the effect of density distribution on SC for different hybridizations.

\subsection{Single-particle Green function and self-energy}

To investigate the metallic or insulating nature upon suppressing the SC phase via interplay between the density distribution and hybridization, we examine the spectral function $A(k,\omega=0)=\mathrm{-\frac{1}{\pi}Im}G(k,\omega=0)$ by linearly extrapolating to the zero imaginary-frequency limit from the two lowest Matsubara frequencies of Im$G(\mathbf{K},iw_n)$ (accurately computed using the DCA method)~\cite{single-band,WeiWu2018, lx2025}. This extrapolated value serves as a reliable proxy for the spectral function $A(k,\omega=0)$ at zero real frequency. 

The convergence of Im$\Sigma(\mathbf{K},i\omega_n)$ to a finite value as $i\omega_n \to 0$ is a typical non-Fermi liquid behavior; while the diverging or decaying behavior indicate the gap opening originating from strong correlation or gap closing separately~\cite{self-energy} at the particular $\mathbf{K}$ point. Here, we focus on the nodal $\mathbf{K=}(\frac{\pi}{2},\frac{\pi}{2})$ and antinodal $\mathbf{K=}(\pi,0)$ points included in our adopted DCA cluster.

\begin{figure*} 
\psfig{figure=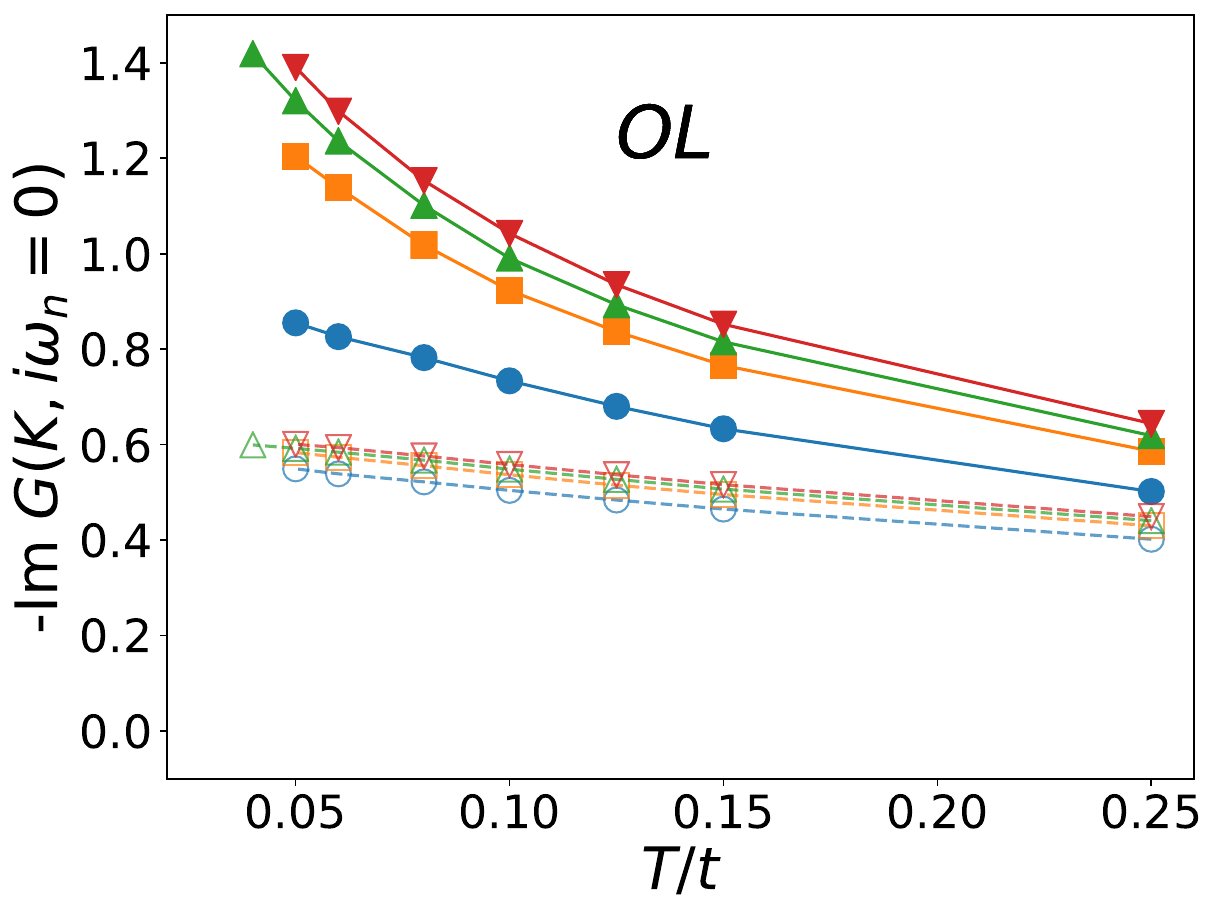, height=6cm,width = .48\textwidth, trim={0 0 0 0}}
\psfig{figure=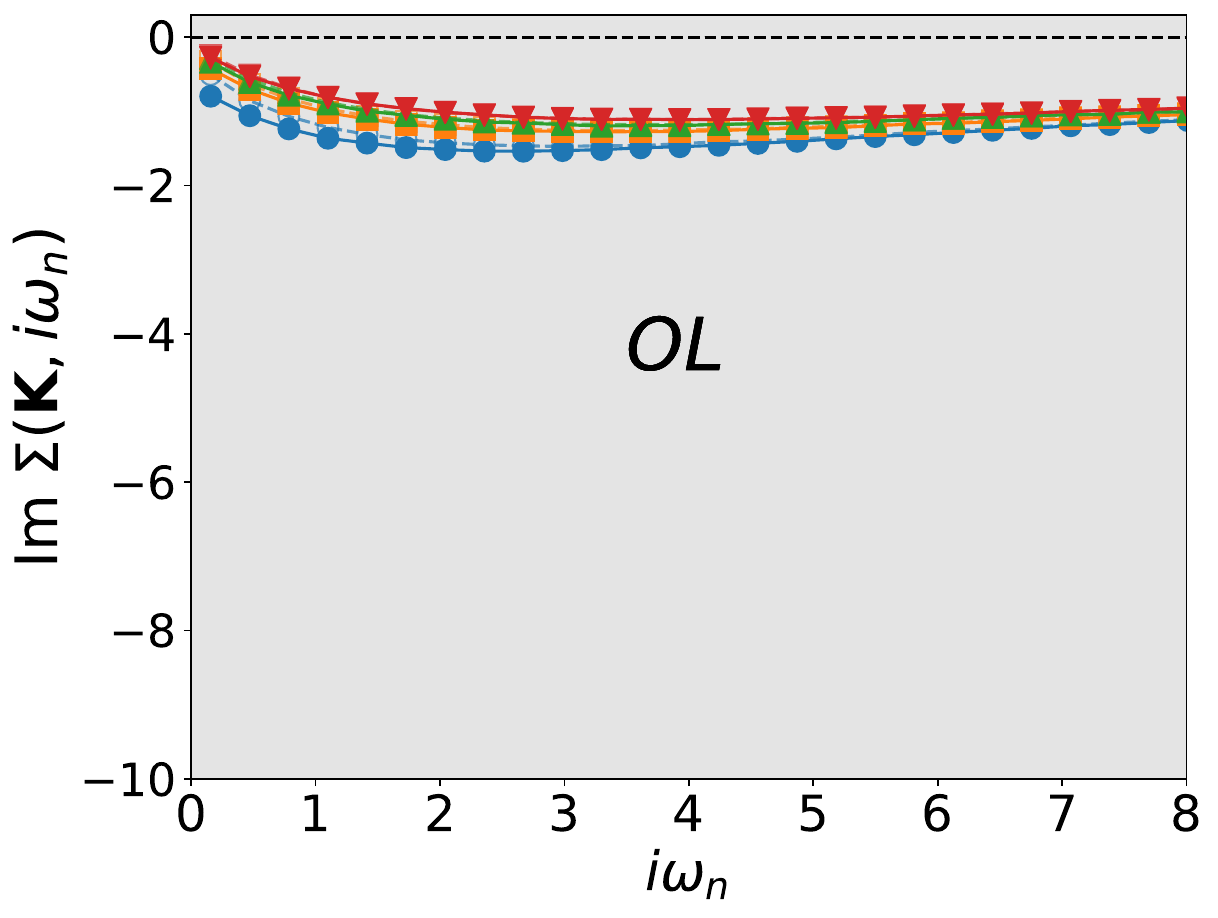, height=6cm,width = .48\textwidth, trim={0 0 0 0}}
\psfig{figure=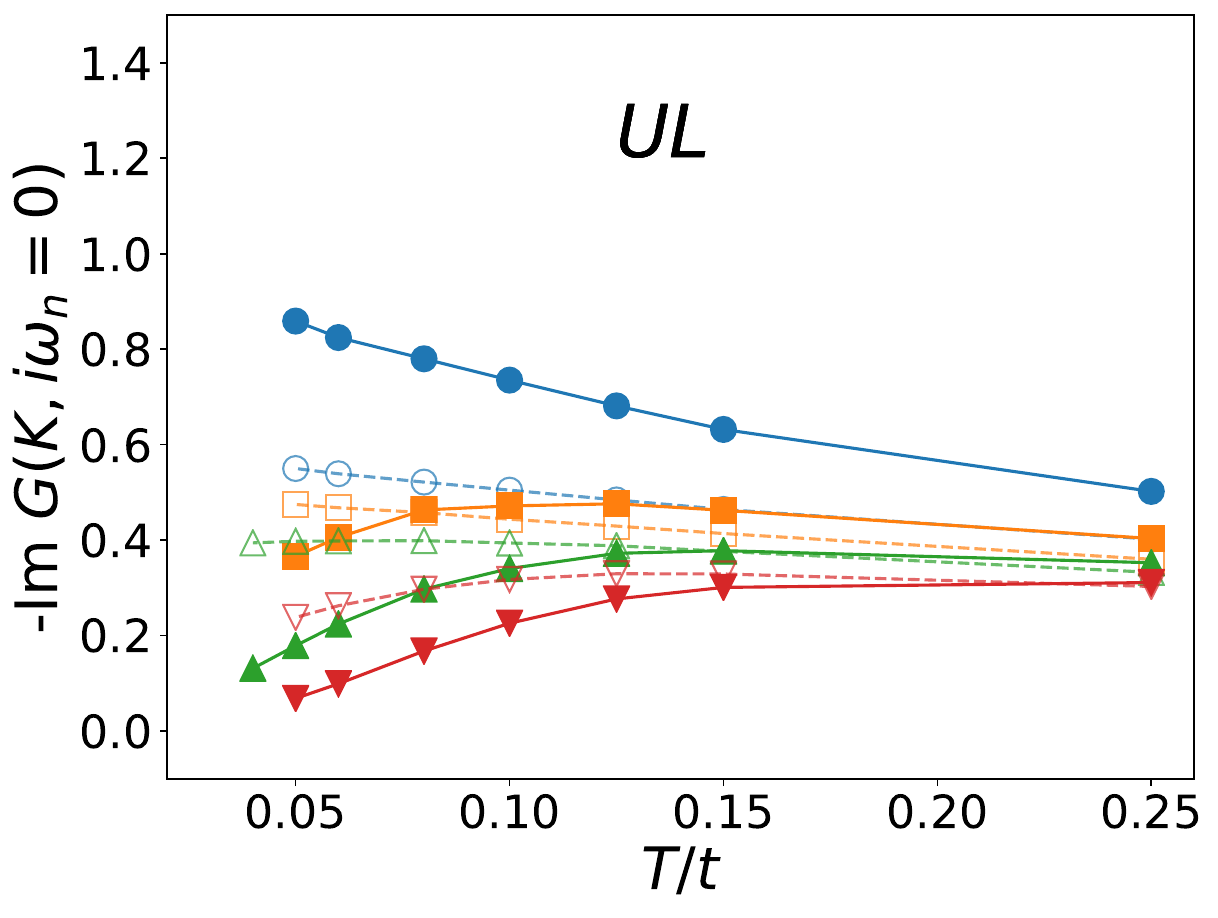, height=6cm,width = .48\textwidth, trim={0 0 0 0}}
\psfig{figure=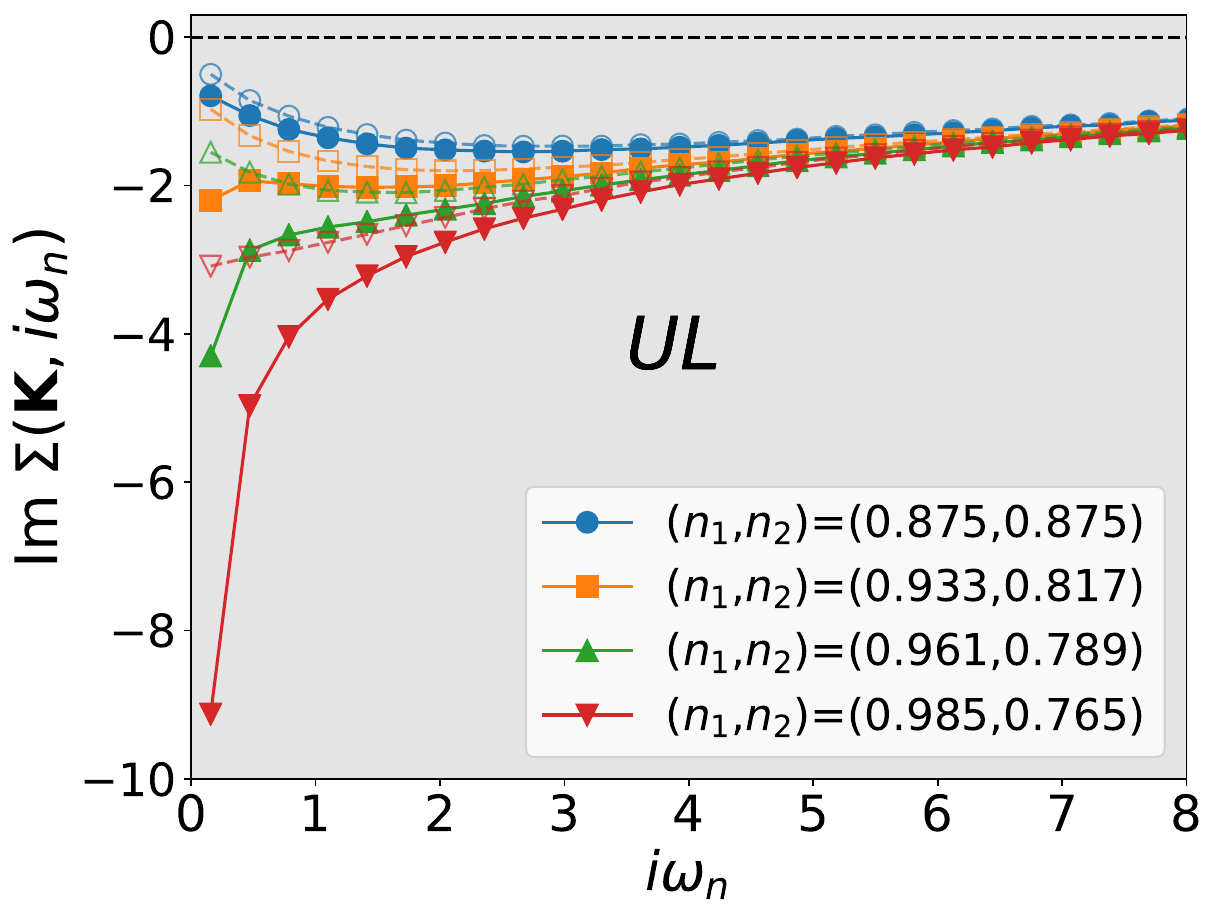, height=6cm,width = .48\textwidth, trim={0 0 0 0}}
\caption{Temperature
evolution of the extrapolated imaginary zero-frequency $-\text{Im}G(\mathbf{K},i\omega_n = 0)$ obtained from a linear extrapolation of the first
two Matsubara frequencies
(white panels) and the imaginary part of
self-energy $\text{Im}\Sigma(\mathbf{K},i\omega_n)$ at $T/t =0.05$ (gray panels) for the OL (upper panels) and UL (lower panels). The dashed and solid lines represent nodal $(\pi/2,\pi/2)$ and antinodal $(\pi,0)$ directions for various density distributions. The hybridization is $t_\perp/t=0.25$ within the $d$-wave pairing regime.}
\label{FL1}
\end{figure*}

As illustrated by the extrapolated single-particle Green function in Fig.~\ref{FL1}, the OL consistently exhibits Fermi liquid behavior across all density distributions at $t_\perp/t=0.25$. In contrast, the UL initially displays Fermi liquid behavior since both nodal and antinodal directions manifest the increasing values with lowering $T$. The UL then develops an energy gap at larger density imbalance, e.g. $(n_1,n_2)=(0.933,0.817)$.
The divergent imaginary part of self-energy observed for the UL at $(n_1,n_2)=(0.933,0.817)$ further corroborates the emergence of strong correlation induced pseudogap opening. The self-energy of OL indicates the strongly renormalized electronic interaction so that metallic character.   

\begin{figure*} 
\psfig{figure=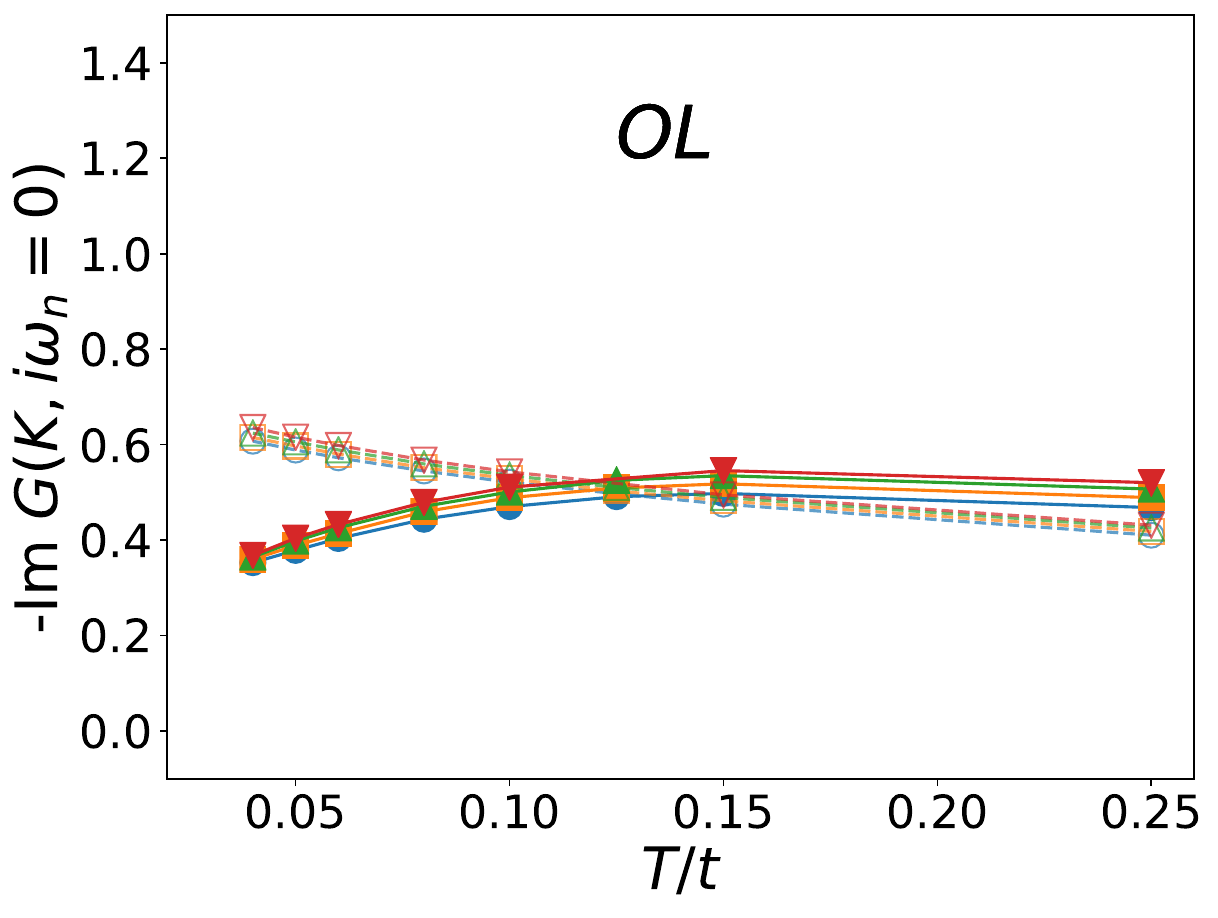, height=6cm,width = .48\textwidth, trim={0 0 0 0}}
\psfig{figure=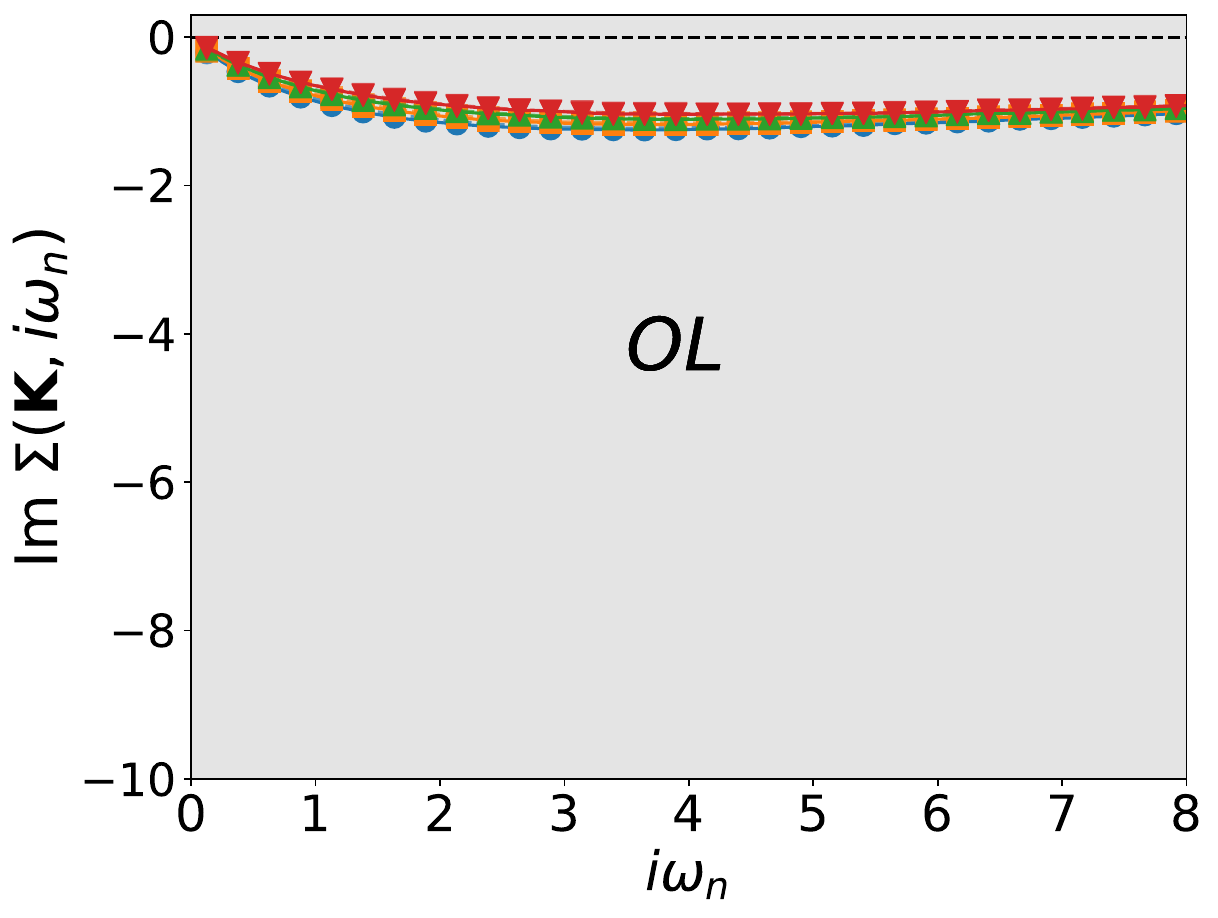, height=6cm,width = .48\textwidth, trim={0 0 0 0}}
\psfig{figure=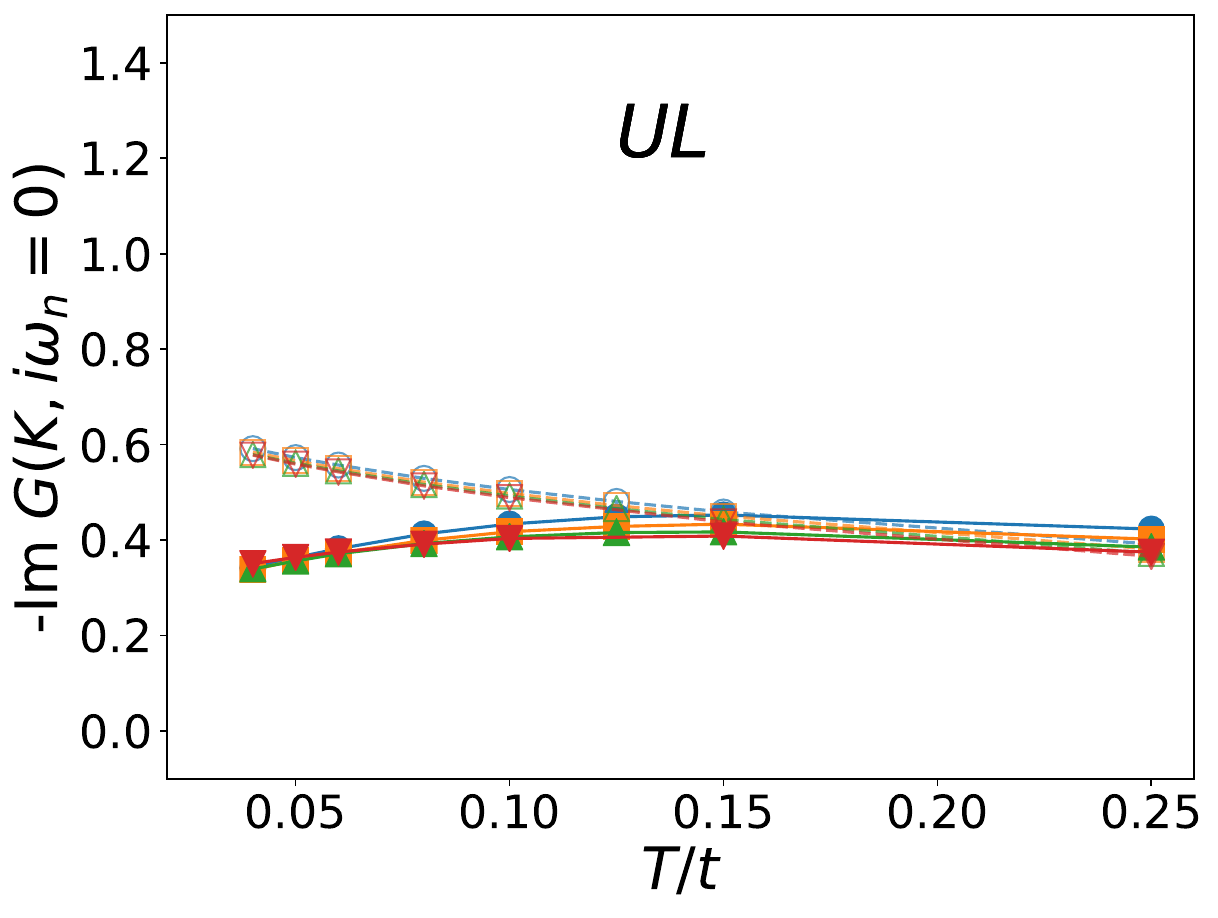, height=6cm,width = .48\textwidth, trim={0 0 0 0}}
\psfig{figure=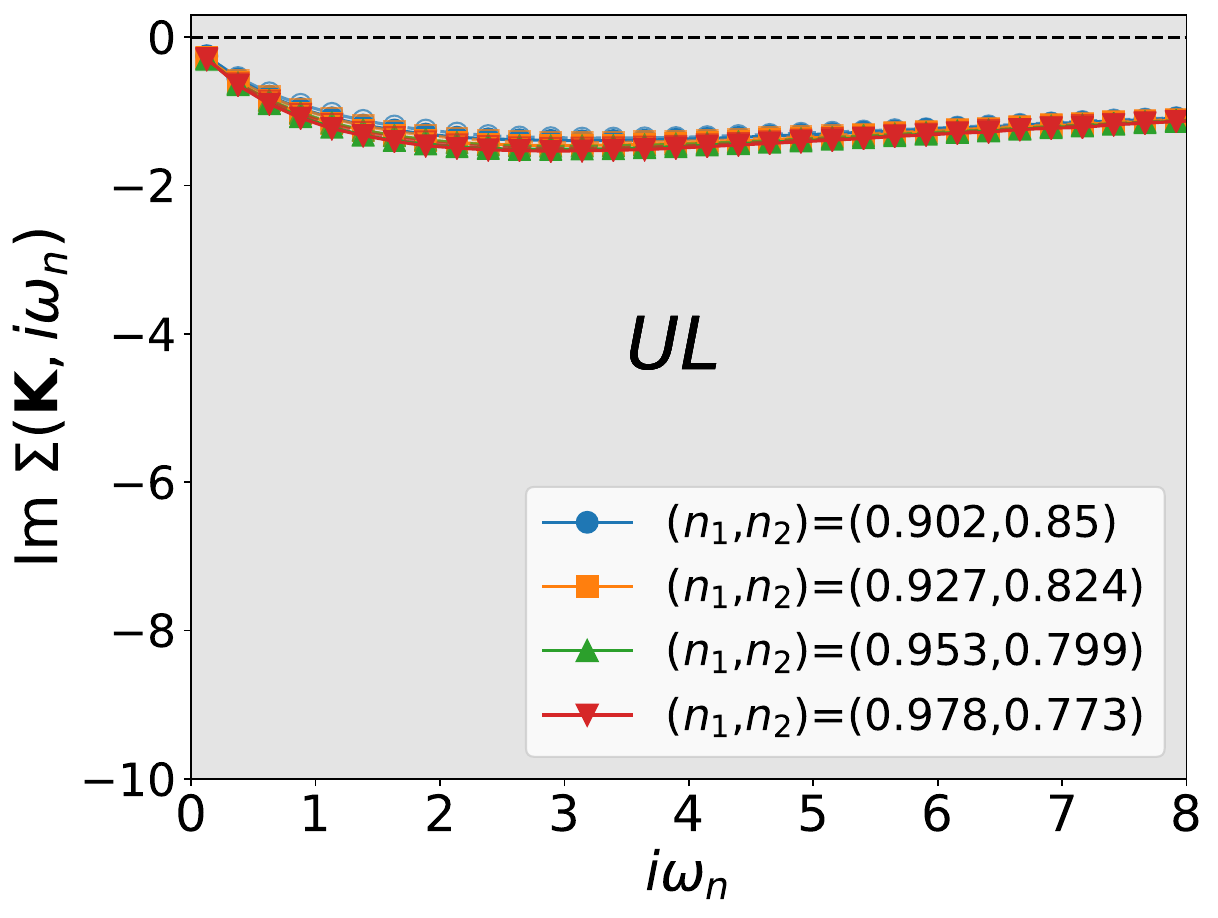, height=6cm,width = .48\textwidth, trim={0 0 0 0}}
\caption{Temperature evolution of the extrapolated imaginary zero-frequency $-\text{Im}G(\mathbf{K},i\omega_n = 0)$ obtained from a linear extrapolation of the first two Matsubara frequencies (white panels) and the imaginary part of self-energy $\text{Im}\Sigma(\mathbf{K},i\omega_n)$ at $T/t =0.04$ (gray panels) for the bonding (upper panels) and anti-bonding band (lower panels). The dashed and solid lines represent nodal $(\pi/2,\pi/2)$ and antinodal $(\pi,0)$ directions for various density distributions. The hybridization is $t_\perp/t=1.0$ within the $s^\pm$-wave pairing regime.}
\label{FL2}
\end{figure*}

Regarding the ultimate electronic state of the $s^\pm$-wave superconducting phase under sufficiently large density disparities (as established in the main text), Fig.~\ref{FL2} demonstrates two key features: (i) the OL and UL both manifest obvious and similar momentum-dependence; (ii) Crucially, the self-energy reveals weakly correlated electron interactions in both OL and UL. These observations collectively suggest a quantum phase transition from the $s^\pm$-wave pairing SC to a Fermi liquid phase beyond the critical density imbalance in Fig.~\ref{phase}.

\clearpage

\bibliography{main}

@article{d_value2,
  title = {Enhanced superconductivity via layer differentiation in the trilayer Hubbard model},
  author = {Liu, Xun and Jiang, Mi},
  journal = {Phys. Rev. B},
  volume = {112},
  issue = {20},
  pages = {L201103},
  numpages = {7},
  year = {2025},
  month = {Nov},
  doi = {10.1103/n28s-ggzc},
  url = {https://link.aps.org/doi/10.1103/n28s-ggzc}
}

@article{cold_2025,
  title = {{Interaction-induced moiré systems in twisted bilayer optical lattices}},
  author = {Jian-Hua Zeng, Qizhong Zhu and Liang He},
 journal = {Phys. Rev. A},
year = {2025},
 url = {https://journals.aps.org/pra/abstract/10.1103/zpqd-ryjm},
 doi = {https://doi.org/10.1103/zpqd-ryjm},
}

@article{bilayer1,
  title = {{Nodeless d-wave pairing in a two-layer Hubbard model}},
  author = {Bulut, Nejat and Scalapino, Douglas J. and Scalettar, Richard T.},
  journal = {Phys. Rev. B},
  volume = {45},
  issue = {10},
  pages = {5577--5584},
  numpages = {0},
  year = {1992},
  month = {Mar},
  publisher = {American Physical Society},
  doi = {10.1103/PhysRevB.45.5577},
  url = {https://link.aps.org/doi/10.1103/PhysRevB.45.5577}
}

@article{bilayer2,
  title = {{Magnetic and pairing correlations in coupled Hubbard planes}},
  author = {Scalettar, Richard T. and Cannon, Joel W. and Scalapino, Douglas J. and Sugar, Robert L.},
  journal = {Phys. Rev. B},
  volume = {50},
  issue = {18},
  pages = {13419--13427},
  numpages = {0},
  year = {1994},
  month = {Nov},
  publisher = {American Physical Society},
  doi = {10.1103/PhysRevB.50.13419},
  url = {https://link.aps.org/doi/10.1103/PhysRevB.50.13419}
}

@article{bilayer3,
  title = {{Magnetism and pairing in Hubbard bilayers}},
  author = {dos Santos, Raimundo R.},
  journal = {Phys. Rev. B},
  volume = {51},
  issue = {21},
  pages = {15540--15546},
  numpages = {0},
  year = {1995},
  month = {Jun},
  publisher = {American Physical Society},
  doi = {10.1103/PhysRevB.51.15540},
  url = {https://link.aps.org/doi/10.1103/PhysRevB.51.15540}
}

@article{bilayer4,
  title={{Competing magnetic orders in a bilayer Hubbard model with ultracold atoms}},
  author={M. Gall and Nicola Wurz and Jens Samland and Chun Fai Chan and Michael K{\"o}hl},
  journal={Nature},
  year={2021},
  volume={589},
  pages={40 - 43},
  url = {https://doi.org/10.1038/s41586-020-03058-x}
}

@article{bilayer5,
  title = {{Superconductivity in the bilayer Hubbard model: Two Fermi surfaces are better than one}},
  author = {Karakuzu, Seher and Johnston, Steven and Maier, Thomas A.},
  journal = {Phys. Rev. B},
  volume = {104},
  issue = {24},
  pages = {245109},
  numpages = {12},
  year = {2021},
  month = {Dec},
  publisher = {American Physical Society},
  doi = {10.1103/PhysRevB.104.245109},
  url = {https://link.aps.org/doi/10.1103/PhysRevB.104.245109}
}

@article{bilayer6,
  title = {{Superfluidity and density order in a bilayer extended Hubbard model}},
  author = {Vanhala, Tuomas I. and Baarsma, Jildou E. and Heikkinen, Miikka O. J. and Troyer, Matthias and Harju, Ari and T\"orm\"a, P\"aivi},
  journal = {Phys. Rev. B},
  volume = {91},
  issue = {14},
  pages = {144510},
  numpages = {8},
  year = {2015},
  month = {Apr},
  publisher = {American Physical Society},
  doi = {10.1103/PhysRevB.91.144510},
  url = {https://link.aps.org/doi/10.1103/PhysRevB.91.144510}
}

@article{bilayer7,
  title = {{Nonlocal density interactions in auxiliary-field quantum Monte Carlo simulations: Application to the square lattice bilayer and honeycomb lattice}},
  author = {Golor, Michael and Wessel, Stefan},
  journal = {Phys. Rev. B},
  volume = {92},
  issue = {19},
  pages = {195154},
  numpages = {11},
  year = {2015},
  month = {Nov},
  publisher = {American Physical Society},
  doi = {10.1103/PhysRevB.92.195154},
  url = {https://link.aps.org/doi/10.1103/PhysRevB.92.195154}
}

@article{maier2011,
  title = {Pair structure and the pairing interaction in a bilayer {H}ubbard model for unconventional superconductivity},
  author = {Maier, T. A. and Scalapino, D. J.},
  journal = {Phys. Rev. B},
  volume = {84},
  issue = {18},
  pages = {180513},
  numpages = {4},
  year = {2011},
  month = {Nov},
  publisher = {American Physical Society},
  doi = {10.1103/PhysRevB.84.180513},
  url = {https://link.aps.org/doi/10.1103/PhysRevB.84.180513}
}

@article{Maier2016,
  title={{$s^{\pm}$ pairing near a Lifshitz transition}},
  author={Vivek Mishra and Douglas J. Scalapino and Thomas A. Maier},
  journal={Scientific Reports},
  year={2016},
  volume={6},
  url={https://doi.org/10.1038/srep32078}
}

@article{lx2025,
  title = {{Hybridization-induced quantum phase transition in the bilayer Hubbard model}},
  author = {Liu, Xun and Jiang, Mi},
  journal = {Phys. Rev. B},
  volume = {111},
  issue = {24},
  pages = {245103},
  numpages = {8},
  year = {2025},
  month = {Jun},
  publisher = {American Physical Society},
  doi = {10.1103/PhysRevB.111.245103},
  url = {https://link.aps.org/doi/10.1103/PhysRevB.111.245103}
}

@article{Maier2022,
  title = {Enhancing ${T}_{\mathrm{c}}$ in a composite superconductor/metal bilayer system: A dynamical cluster approximation study},
  author = {Dee, Philip M. and Johnston, Steven and Maier, Thomas A.},
  journal = {Phys. Rev. B},
  volume = {105},
  issue = {21},
  pages = {214502},
  numpages = {7},
  year = {2022},
  month = {Jun},
  publisher = {American Physical Society},
  doi = {10.1103/PhysRevB.105.214502},
  url = {https://link.aps.org/doi/10.1103/PhysRevB.105.214502}
}

@article{Maier05,
  title = {Quantum cluster theories},
  author = {Maier, Thomas and Jarrell, Mark and Pruschke, Thomas and Hettler, Matthias H.},
  journal = {Rev. Mod. Phys.},
  volume = {77},
  issue = {3},
  pages = {1027--1080},
  numpages = {0},
  year = {2005},
  month = {Oct},
  publisher = {American Physical Society},
  doi = {10.1103/RevModPhys.77.1027},
  url = {https://link.aps.org/doi/10.1103/RevModPhys.77.1027}
}

@article{GullCTAUX,
doi = {10.1209/0295-5075/82/57003},
url = {https://dx.doi.org/10.1209/0295-5075/82/57003},
year = {2008},
month = {may},
publisher = {},
volume = {82},
number = {5},
pages = {57003},
author = {Gull, E. and Werner, P. and Parcollet, O. and Troyer, M.},
title = {Continuous-time auxiliary-field Monte Carlo for quantum impurity models},
journal = {Europhysics Letters},
}

@inbook{Scalapino2007numerical,
  author    = {Scalapino, D. J.},
  title     = {Numerical Studies of the {2D} Hubbard Model},
  booktitle = {Handbook of High-Temperature Superconductivity: Theory and Experiment},
  year      = {2007},
  editor    = {Schrieffer, J. Robert and Brooks, James S.}, 
  publisher = {Springer},
  address   = {New York, NY},
  pages     = {495--526},
  doi       = {10.1007/978-0-387-68734-6_13},
  url       = {https://doi.org/10.1007/978-0-387-68734-6_13}
}

@article{Maier2006,
  title = {Structure of the Pairing Interaction in the Two-Dimensional {H}ubbard Model},
  author = {Maier, T. A. and Jarrell, M. S. and Scalapino, D. J.},
  journal = {Phys. Rev. Lett.},
  volume = {96},
  issue = {4},
  pages = {047005},
  numpages = {4},
  year = {2006},
  month = {Feb},
  publisher = {American Physical Society},
  doi = {10.1103/PhysRevLett.96.047005},
  url = {https://link.aps.org/doi/10.1103/PhysRevLett.96.047005}
}

@article{Hettler98,
  title = {Nonlocal dynamical correlations of strongly interacting electron systems},
  author = {Hettler, M. H. and Tahvildar-Zadeh, A. N. and Jarrell, M. and Pruschke, T. and Krishnamurthy, H. R.},
  journal = {Phys. Rev. B},
  volume = {58},
  issue = {12},
  pages = {R7475--R7479},
  numpages = {0},
  year = {1998},
  month = {Sep},
  publisher = {American Physical Society},
  doi = {10.1103/PhysRevB.58.R7475},
  url = {https://link.aps.org/doi/10.1103/PhysRevB.58.R7475}
}

@article{code,
title = {{DCA}++: A software framework to solve correlated electron problems with modern quantum cluster methods},
journal = {Computer Physics Communications},
volume = {246},
pages = {106709},
year = {2020},
issn = {0010-4655},
doi = {https://doi.org/10.1016/j.cpc.2019.01.006},
url = {https://www.sciencedirect.com/science/article/pii/S0010465519300086},
author = {Urs R. Hähner and Gonzalo Alvarez and Thomas A. Maier and Raffaele Solcà and Peter Staar and Michael S. Summers and Thomas C. Schulthess},
}

@article{327,
  title={{Signatures of superconductivity near 80 K in a nickelate under high pressure}},
  author={Sun, Hualei and Huo, Mengwu and Hu, Xunwu and Li, Jingyuan and Liu, Zengjia and Han, Yifeng and Tang, Lingyun and Mao, Zhongquan and Yang, Pengtao and Wang, Bosen and others},
  journal={Nature},
  volume={621},
  number={7979},
  pages={493--498},
  year={2023},
  publisher={Nature Publishing Group UK London},
  url= {https://doi.org/10.1038/s41586-023-06408-7}
}

@article{single-band,
  title = {Momentum-space anisotropy and pseudogaps: A comparative cluster dynamical mean-field analysis of the doping-driven metal-insulator transition in the two-dimensional Hubbard model},
  author = {Gull, E. and Ferrero, M. and Parcollet, O. and Georges, A. and Millis, A. J.},
  journal = {Phys. Rev. B},
  volume = {82},
  issue = {15},
  pages = {155101},
  numpages = {14},
  year = {2010},
  month = {Oct},
  publisher = {American Physical Society},
  doi = {10.1103/PhysRevB.82.155101},
  url = {https://link.aps.org/doi/10.1103/PhysRevB.82.155101}
}

@article{WeiWu2018,
  title = {{Pseudogap and Fermi-Surface Topology in the Two-Dimensional Hubbard Model}},
  author = {Wu, Wei and Scheurer, Mathias S. and Chatterjee, Shubhayu and Sachdev, Subir and Georges, Antoine and Ferrero, Michel},
  journal = {Phys. Rev. X},
  volume = {8},
  issue = {2},
  pages = {021048},
  numpages = {17},
  year = {2018},
  month = {May},
  publisher = {American Physical Society},
  doi = {10.1103/PhysRevX.8.021048},
  url = {https://link.aps.org/doi/10.1103/PhysRevX.8.021048}
}

@article{Maier2019,
  title={{Pairfield fluctuations of a 2D Hubbard model}},
  author={Thomas A. Maier and Douglas J. Scalapino},
  journal={npj Quantum Materials},
  year={2019},
  volume={4},
  pages={1-4},
  url={https://doi.org/10.1038/s41535-019-0169-9}
}

@article{Kivelson,
title = {Making high {$T_c$} higher: a theoretical proposal},
author = {Kivelson, Steven A.},
journal = {Physica B: Condensed Matter},
volume = {318},
number = {1},
pages = {61-67},
year = {2002},
issn = {0921-4526},
doi = {https://doi.org/10.1016/S0921-4526(02)00775-5},
url = {https://www.sciencedirect.com/science/article/pii/S0921452602007755}
}

@article{Kivelson2,
  title = {Route to high-temperature superconductivity in composite systems},
  author = {Berg, Erez and Orgad, Dror and Kivelson, Steven A.},
  journal = {Phys. Rev. B},
  volume = {78},
  issue = {9},
  pages = {094509},
  numpages = {7},
  year = {2008},
  month = {Sep},
  publisher = {American Physical Society},
  doi = {10.1103/PhysRevB.78.094509},
  url = {https://link.aps.org/doi/10.1103/PhysRevB.78.094509}
}

@article{Dror,
  title = {Superfluid stiffness renormalization and critical temperature enhancement in a composite superconductor},
  author = {Wachtel, Gideon and Bar-Yaacov, Assaf and Orgad, Dror},
  journal = {Phys. Rev. B},
  volume = {86},
  issue = {13},
  pages = {134531},
  numpages = {11},
  year = {2012},
  month = {Oct},
  publisher = {American Physical Society},
  doi = {10.1103/PhysRevB.86.134531},
  url = {https://link.aps.org/doi/10.1103/PhysRevB.86.134531}
}

@article{maier2008,
  title = {{Enhanced Superconductivity in Superlattices of High-${T}_{c}$ Cuprates}},
  author = {Okamoto, Satoshi and Maier, Thomas A.},
  journal = {Phys. Rev. Lett.},
  volume = {101},
  issue = {15},
  pages = {156401},
  numpages = {4},
  year = {2008},
  month = {Oct},
  publisher = {American Physical Society},
  doi = {10.1103/PhysRevLett.101.156401},
  url = {https://link.aps.org/doi/10.1103/PhysRevLett.101.156401}
}

@article{fontenele_increasing_2024,
	title = {Increasing superconducting ${T}_{c}$ by layering in the attractive {H}ubbard model},
        author = {Fontenele, Rodrigo A. and Costa, Natanael C. and Paiva, Thereza and dos Santos, Raimundo R.},
        journal = {Phys. Rev. A},
	volume = {110},
	number = {5},
        pages = {053315},
	year = {2024},
	url = {https://link.aps.org/doi/10.1103/PhysRevA.110.053315},
	doi = {10.1103/PhysRevA.110.053315},
}

@article{eisenstein_2004,
        title={Bose–Einstein condensation of excitons in bilayer electron systems},
        author={James Philip Eisenstein and Allan H Macdonald},
        journal={Nature},
        year={2004},
        volume={432},
        pages={691-694},
        url={https://doi.org/10.1038/nature03081},
        doi = {10.1038/nature03081}
}

@article{rademaker_2013,
        title={Determinant quantum Monte Carlo study of exciton condensation in the bilayer {H}ubbard model},
        author={Rademaker, Louk and Johnston, Steve and  Zaanen, Jan and Jeroen, Van Den Brink},
        journal={Phys. Rev. B},
        volume={88},
        number={23},
        pages = {235115},
        year={2013},
	url = {https://link.aps.org/doi/10.1103/PhysRevB.88.235115},
}

@article{huang_biexciton_2020,
        title={Biexciton Condensation in Electron-Hole-Doped {H}ubbard Bilayers: A Sign-Problem-Free Quantum Monte Carlo Study},
        author={ Huang, Xu Xin  and  Claassen, Martin  and  Huang, Edwin W.  and  Moritz, Brian  and  Devereaux, Thomas P. },
        journal={Phys. Rev. Lett},
        volume={124},
        number={7},
        pages = {077601},
        year={2020},
	url = {https://link.aps.org/doi/10.1103/PhysRevLett.124.077601},
}

@article{1999magnetic,
        title={Magnetic fluctuations in coupled inequivalent {H}ubbard layers as a model for {Y}$_2${Ba}$_4${Cu}$_7${O}$_{15}$},
        author={Hildebrand, G and Arrigoni, E and Hanke, W and Schmalian, J},
        journal={European Physical Journal. B},
        volume={8},
        pages={195–205},
        year={1999},
        doi= {https://doi.org/10.1007/s100510050682},
        url = {https://link.springer.com/article/10.1007/s100510050682},
}

@article{euverte_magnetic_2013,
        title={Magnetic transition in a correlated band insulator},
        author={ Euverte, A.  and  Chiesa, S.  and  Scalettar, R. T.  and  Batrouni, G. G. },
        journal={Phys. Rev. B},
        volume={87},
        number={12},
        pages={269-275},
        year={2013},
	url = {https://link.aps.org/doi/10.1103/PhysRevB.87.125141},
	doi = {10.1103/PhysRevB.87.125141},
}

@article{ruger_phase_2014,
	title = {{The phase diagram of the square lattice bilayer Hubbard model: a variational Monte Carlo study}},
        author = {Rüger, Robert and Tocchio, Luca F and Valentí, Roser and Gros, Claudius},
        journal = {New Journal of Physics}, 
	volume = {16},
        number = {3},
        pages = {033010},
        year={2014},
	url = {https://dx.doi.org/10.1088/1367-2630/16/3/033010},
	doi = {10.1088/1367-2630/16/3/033010},
}

@article{lin_magnetic_2024,
        title={Magnetic phase diagram of a two-orbital model for bilayer nickelates with varying doping},
        author={ Lin, Ling Fang  and  Zhang, Yang  and  Kaushal, Nitin  and  Alvarez, Gonzalo  and  Maier, Thomas A.  and  Moreo, Adriana  and  Dagotto, Elbio },
        journal={Phys. Rev. B},
        volume = {110},
        number={19},
        pages={195135},
        year={2024},
        url = {https://link.aps.org/doi/10.1103/PhysRevB.110.195135},
        doi = {10.1103/PhysRevB.110.195135},
}

@article{zhang_optimizing_2025,
	title = {Optimizing the critical temperature and superfluid density of a metal-superconductor bilayer},
	author = {Zhang, Yutan and Dee, Philip M. and Cohen-Stead, Benjamin and Maier, Thomas       A. and Johnston, Steven and Scalettar, Richard},
        journal={Phys. Rev. B},
	volume = {112},
        number = {6},
	pages = {064510},
	year = {2025},
	url = {https://link.aps.org/doi/10.1103/lcgr-bqcv},
	doi = {10.1103/lcgr-bqcv},
}

@article{pan_competition_2024,
	title = {Competition between $d$-wave and $d+is$-wave superconductivity in the {H}ubbard model on a checkerboard lattice},
	author = {Pan, Yue and Ma, Runyu and Chen, Chao and Jia, Zixuan and Ma, Tianxing},
        journal = {Phys. Rev. B},
	volume = {110},
        number = {14},
        pages = {144509},
        year = {2024},
	url = {https://link.aps.org/doi/10.1103/PhysRevB.110.144509},
	doi = {10.1103/PhysRevB.110.144509},
}

@article{ZYY,
  title = {${s}^{\ifmmode\pm\else\textpm\fi{}}$-wave superconductivity in the bilayer two-orbital {H}ubbard model},
  author = {Zheng, Yao-Yuan and W\'u, W\'ei},
  journal = {Phys. Rev. B},
  volume = {111},
  issue = {3},
  pages = {035108},
  numpages = {7},
  year = {2025},
  month = {Jan},
  publisher = {American Physical Society},
  doi = {10.1103/PhysRevB.111.035108},
  url = {https://link.aps.org/doi/10.1103/PhysRevB.111.035108}
}

@article{LJX,
  title = {{Transition from $s_{\pm}$-wave to $d_{{x}^{2}-{y}^{2}}$-wave superconductivity driven by interlayer interaction in the bilayer two-orbital model of La$_3$Ni$_2$O$_7$}},
  author = {Xi, Wenhan and Yu, Shun-Li and Li, Jian-Xin},
  journal = {Phys. Rev. B},
  volume = {111},
  issue = {10},
  pages = {104505},
  numpages = {10},
  year = {2025},
  month = {Mar},
  publisher = {American Physical Society},
  doi = {10.1103/PhysRevB.111.104505},
  url = {https://link.aps.org/doi/10.1103/PhysRevB.111.104505}
}

@Article{wang_2024,
title = {{Normal and Superconducting Properties of La$_3$Ni$_2$O$_7$}},
author = {Meng Wang and Hai-Hu Wen and Tao Wu and Dao-Xin Yao and Tao Xiang},
journal = {Chin. Phys. Lett.},
volume = {41},
number = {7},
pages = {077402},
year = {2024},
doi = {10.1088/0256-307X/41/7/077402},
url = {http://cpl.iphy.ac.cn/en/article/doi/10.1088/0256-307X/41/7/077402}
}

@article{yang_2024,
title = {Orbital-dependent electron correlation in double-layer nickelate {La}$_3${Ni}$_2${O}$_7$},
author = {Yang, Jiangang and Sun, Hualei and Hu, Xunwu and Xie, Yuyang and Miao, Taimin and Luo, Hailan and Chen, Hao and Liang, Bo and Zhu, Wenpei and Qu, Gexing and Chen, Cui-Qun and Huo, Mengwu and Huang, Yaobo and Zhang, Shenjin and Zhang, Fengfeng and Yang, Feng and Wang, Zhimin and Peng, Qinjun and Mao, Hanqing and Liu, Guodong and Xu, Zuyan and Qian, Tian and Yao, Dao-Xin and Wang, Meng and Zhao, Lin and Zhou, X. J.},
journal = {Nature Communications},
volume = {15},
number = {1},
pages = {4373},
year = {2024},
url = {https://doi.org/10.1038/s41467-024-48701-7},
doi = {10.1038/s41467-024-48701-7},
}

@misc{maier327,
title={{Interlayer Pairing in Bilayer Nickelates}}, 
author={Thomas A. Maier and Peter Doak and Ling-Fang Lin and Yang Zhang and Adriana Moreo and Elbio Dagotto},
year={2025},
eprint={2506.07741},
archivePrefix={arXiv},
}

@misc{cold2,
title={Thermodynamics and density fluctuations in a bilayer {H}ubbard system of ultracold atoms}, 
author={J. Samland and N. Wurz and M. Gall and M. Köhl},
year={2024},
eprint={2407.11863},
archivePrefix={arXiv},
}

@article{cold3,
title={Observation of a bilayer superfluid with interlayer coherence}, 
author={Erik, Rydow and Vijay Pal Singh and  Abel, Beregi and  En Chang and Ludwig, Mathey and  Christopher, J. Foot and Shinichi Sunami},
journal = {Nature Communications},
year={2025},
pages = {7201},
url = {https://doi.org/10.1038/s41467-025-62277-w},
doi = {https://doi.org/10.1038/s41467-025-62277-w},
}

@article{bohrdt_2021,
title = {Exploration of doped quantum magnets with ultracold atoms},
author = {Bohrdt, Annabelle and Homeier, Lukas and Reinmoser, Christian and Demler, Eugene and Grusdt, Fabian},
journal = {Annals of Physics},
volume = {435},
pages = {168651},
year = {2021},
url = {https://www.sciencedirect.com/science/article/pii/S0003491621002578},
doi = {https://doi.org/10.1016/j.aop.2021.168651},
}

@article{yang_possible_2023,
	title = {Possible s$_\pm$-wave superconductivity in {La}$_3${Ni}$_2${O}$_7$},
	author = {Yang, Qing-Geng and Wang, Da and Wang, Qiang-Hua},
	journal = {Phys. Rev. B},
	volume = {108},
	number = {14},
	pages = {L140505},
	year = {2023},
	url = {https://link.aps.org/doi/10.1103/PhysRevB.108.L140505},
	doi = {10.1103/PhysRevB.108.L140505},
}

@article{lu_2024,
	title = {Interlayer-Coupling-Driven High-Temperature Superconductivity in {La}$_3${Ni}$_2${O}$_7$ under Pressure},
	author = {Lu, Chen and Pan, Zhiming and Yang, Fan and Wu, Congjun},
	journal = {Phys. Rev. Lett.},
	volume = {132},
	number = {14},
	pages = {146002},
	year = {2024},
	url = {https://link.aps.org/doi/10.1103/PhysRevLett.132.146002},
	doi = {10.1103/PhysRevLett.132.146002},
}

@article{zhang_trends_2023,
	title = {Trends in electronic structures and $s_{\pm}$-wave pairing for the rare-earth series in bilayer nickelate superconductor {R}$_3${Ni}$_2${O}$_7$},
	author = {Zhang, Yang and Lin, Ling-Fang and Moreo, Adriana and Maier, Thomas A. and Dagotto, Elbio},
	journal = {Phys. Rev. B},
	volume = {108},
	number = {16},
	pages = {165141},
	year = {2023},
	url = {https://link.aps.org/doi/10.1103/PhysRevB.108.165141},
	doi = {10.1103/PhysRevB.108.165141},
}

@article{botzel_theory_2024,
	title = {Theory of magnetic excitations in the multilayer nickelate superconductor {La}$_3${Ni}$_2${O}$_7$},
	author = {Bötzel, Steffen and Lechermann, Frank and Gondolf, Jannik and Eremin, Ilya M.},
	journal = {Phys. Rev. B},
	volume = {109},
	number = {18},
	pages = {L180502},
	year = {2024},
	url = {https://link.aps.org/doi/10.1103/PhysRevB.109.L180502},
	doi = {10.1103/PhysRevB.109.L180502},
}

@article{liao_electron_2023,
	title = {Electron correlations and superconductivity in {La}$_3${Ni}$_2${O}$_7$ under pressure tuning},
	author = {Liao, Zhiguang and Chen, Lei and Duan, Guijing and Wang, Yiming and Liu, Changle and Yu, Rong and Si, Qimiao},
	journal = {Phys. Rev. B},
	volume = {108},
	number = {21},
	pages = {214522},
	year = {2023},
	url = {https://link.aps.org/doi/10.1103/PhysRevB.108.214522},
	doi = {10.1103/PhysRevB.108.214522},
}

@article{luo_2024,
	title = {High-{TC} superconductivity in {La}$_3${Ni}$_2${O}$_7$ based on the bilayer two-orbital {t}-{J} model},
	author = {Luo, Zhihui and Lv, Biao and Wang, Meng and Wú, Wéi and Yao, Dao-Xin},
	journal = {npj Quantum Materials},
	volume = {9},
	number = {1},
	pages = {61},
	year = {2024},
	url = {https://doi.org/10.1038/s41535-024-00668-w},
	doi = {10.1038/s41535-024-00668-w},
	date = {2024-08-13},
}

@article{botana_2024,
	title = {Polymorphism in the Ruddlesden–Popper Nickelate {La}$_3${Ni}$_2${O}$_7$: Discovery of a Hidden Phase with Distinctive Layer Stacking},
	author = {Chen, Xinglong and Zhang, Junjie and Thind, Arashdeep S. and Sharma, Shekhar and {LaBollita}, Harrison and Peterson, Gordon and Zheng, Hong and Phelan, Daniel P. and Botana, Antia S. and Klie, Robert F. and Mitchell, J. F.},
	journal = {J. Am. Chem. Soc.},
	volume = {146},
	number = {6},
	pages = {3640--3645},
	year = {2024},
	url = {https://doi.org/10.1021/jacs.3c14052},
	doi = {10.1021/jacs.3c14052},
}

@article{self-energy,
  title = {Competition between band and Mott insulators in the bilayer Hubbard model: A dynamical cluster approximation study},
  author = {Lee, Hunpyo and Zhang, Yu-Zhong and Jeschke, Harald O. and Valent\'{\i}, Roser},
  journal = {Phys. Rev. B},
  volume = {89},
  issue = {3},
  pages = {035139},
  numpages = {6},
  year = {2014},
  month = {Jan},
  publisher = {American Physical Society},
  doi = {10.1103/PhysRevB.89.035139},
  url = {https://link.aps.org/doi/10.1103/PhysRevB.89.035139}
}

@article{2024Dissipationless,
  title={Dissipationless Counterflow Currents above ${T}_c$ in Bilayer Superconductors},
  author={ Homann, Guido  and  Michael, Marios H.  and  Cosme, Jayson G.  and  Mathey, Ludwig },
  journal={Phys. Rev. Lett.},
  volume={132},
  number={9},
  pages={6},
  year={2024},
  doi = {10.1103/PhysRevLett.132.096002},
  url = {https://link.aps.org/doi/10.1103/PhysRevLett.132.096002}
}

@article{eigenlog,
  title = {{Pair Tunneling as a Probe of Fluctuations in Superconductors}},
  author = {Scalapino, D. J.},
  journal = {Phys. Rev. Lett.},
  volume = {24},
  issue = {19},
  pages = {1052--1055},
  numpages = {0},
  year = {1970},
  month = {May},
  publisher = {American Physical Society},
  doi = {10.1103/PhysRevLett.24.1052},
  url = {https://link.aps.org/doi/10.1103/PhysRevLett.24.1052}
}

\end{document}